\documentclass{aa}  

\usepackage{graphicx}
\usepackage{txfonts}
\usepackage{xcolor}
\usepackage{siunitx}
\usepackage{comment}
\usepackage{pdflscape} 
\usepackage{subcaption}

\begin{document} 

   \title{Identification of Large Equivalent Width Dusty Galaxies at 4 $<$ z $<$ 6 from Sub-mm Colours}
   \titlerunning{}


   \author{D. Burgarella
          \inst{1}
          \and
          P. Theul\'e
          \inst{1}
          \and
          V. Buat
          \inst{1}
          \and
          L. Gouiran
          \inst{1}
          L. Turco
          \inst{1}
          M. Boquien
          \inst{2}
          \and
          T. J. L. C. Bakx
          \inst{3, 4}
          \and
          A. K. Inoue
          \inst{5, 6}
          \and
          Y. Fudamoto
          \inst{4}
          \and
          Y. Sugahara
          \inst{4, 6}
          \and
          J. Zavala
          \inst{4}
          }

 \institute{
   Aix Marseille Univ, CNRS, CNES, LAM, Marseille, France\\
   \email{denis.burgarella@lam.fr}
 \and
   Centro de Astronomía (CITEVA), Universidad de Antofagasta, Avenida Angamos 601, Antofagasta, Chile
 \and
   Division of Particle and Astrophysical Science, Graduate School of Science, Nagoya University, Aichi 464-8602, Japan
 \and
   National Astronomical Observatory of Japan, 2-21-1, Osawa, Mitaka, Tokyo 181-8588, Japan
 \and
   Department of Physics, School of Advanced Science and Engineering, Faculty of Science and Engineering, Waseda University, 3-4-1, Okubo, Shinjuku, Tokyo 169-8555
 \and
   Waseda Research Institute for Science and Engineering, Faculty of Science and Engineering, Waseda University, 3-4-1, Okubo, Shinjuku, Tokyo 169-8555
      }
   \date{Received May 8, 2022; accepted March 16, 1997}

 
  \abstract
   {Infrared (IR), sub-millimetre (sub-mm) and millimetre (mm) databases contain a huge quantity of high quality data. However, a large part of these data are photometric, and are thought not to be useful to derive a quantitative information on the nebular emission of galaxies.}
   {The aim of this project is first to identify galaxies at z $\gtrsim$ 4-6, and in the epoch of reionization from their sub-mm colours. We also aim at  showing that the colours can be used to try and derive physical constraints from photometric bands, when accounting for the contribution from the IR fine structure lines to these photometric bands.}
   {We model the flux of IR fine structure lines with CLOUDY, and add them to the dust continuum emission with CIGALE. Including or not emission lines in the simulated spectral energy distribution (SED) modifies the broad band emission and colours.}
   {The introduction of the lines allows to identify strong star forming galaxies at z $\gtrsim$ 4 - 6  from the [$\log_{10} \frac {PSW_{250 \si{\um}}} {PMW_{350 \si{\um}}}$ versus $\log_{10} \frac {LABOCA_{870 \si{\um}}} {PLW_{500 \si{\um}}}$] colour-colour diagramme. By comparing the relevant models to each observed galaxy colour, we are able to roughly estimate the fluxes of the lines, and the associated nebular parameters. This method allows to identify a double sequence in a plot built from the ionization parameter and the gas metallicity.}
   {The HII and photodissociation region (PDR) fine structure lines are an essential part of the SEDs. It is important to add them when modelling the spectra, especially at z $\gtrsim$ 4 - 6 where their equivalent widths can be large. Conversely, we show that we can extract some information on strong IR fine structure lines and on the physical parameters related to the nebular emission from IR colour-colour diagrams.}


   \maketitle
%
\section{Introduction}

Several papers reported excesses of the flux densities of high redshift galaxies in broad bands. For instance, a boost of the Spitzer/IRAC bands at z $\sim$ 7 - 8 is observed when H$\alpha$, and [OIII]500.7 nm fall in the mid-infrared (mid-IR) filters \citep[e.g., ][]{deBarros2013, Roberts-Borsani2020, Anders2003}. In the sub-millimetre (sub-mm) as well, \cite{Seaquist2004} suggested that about 25~\% of the 850~\si{\um} flux density could be due to the CO(3--2) molecular emission. \
More relevant to this paper, \cite{Smail2011} estimated that a [CII]158 \si{\um} fine structure line with 0.27~\% of the galaxy’s L$_{dust}$ would contribute 5 - 10~\% to the far-IR broadband flux densities at 850 \si{\um}. Because the line contribution scales linearly with  L$_{[CII]} / L_{dust}$, the [CII]158 \si{\um} to dust continuum luminosity ratio, sources with L$_{[CII]} / L_{dust}$ > 1~\% will contribute more than four time this amount, reaching 20 to 40~\% of the 850 \si{\um} flux densities for galaxies at z $\sim$ 4 - 6 (also see \cite{Seymour2012} for the contribution to the Herschel/SPIRE 500 \si{\um} band). 

On the contrary, L$_{[CII]}/ L_{dust}$ presents a deficit for galaxies with large IR surface brightnesses or IR luminosities. \cite{Luhman2003} proposed that this deficit could be due to high values of the ionization parameter\footnote{The ionization parameter is defined as the dimensionless ratio of the incident ionizing photon density to the hydrogen density: $U = Q(H) / (4 \pi R^2 n_H c)$ where Q(H) is the number of hydrogen ionizing photons per second, c is the speed of light, n$_H$ is the hydrogen density, and R is the distance of the ionizing source to the illuminated face. } ($\log_{10}$~U~>~-2.5), for which a narrower photodissociation region (PDR) would lead to lower [CII]158~\si{\um} fluxes. This explanation is also supported by a number of other analyses \citep[e.g., ][]{Abel2009, Diaz-Santos2013, Diaz-Santos2017, Herrera-Camus2018}.

Recent promising papers from the James Webb Space Telescope (JWST) \citep[e.g.,][]{Schaerer2022, Trump2022, Taylor2022} suggest that we are now able to spectroscopically measure the nebular parameters of galaxies in the epoch of reionization (EoR) at z $\sim$ 5 - 8.  However, we still have large uncertainties when estimating the obscured star formation density at high redshift \citep[e.g., ][]{Algera2022}. Thus, we still need to identify and measure the luminosities of galaxies in the EoR to better understand how the total star formation rate density (SFRD) evolves in the early Universe to understand the formation and early evolution of galaxies at z $>$ 10 \citep[e.g., ][]{Finkelstein2022}.

In this paper, we propose an original method to identify 4.5~$<$~z~$<$~6.0 galaxies via a broad-band  excess due to the  [CII]158~\si{\um} line. This method is applicable to very large numbers of dusty galaxies that can be extracted from the already existing far-IR and sub-mm databases (Section~3). We present tests on a sample of galaxies with spectroscopic redshifts observed with the South Pole Telescope (SPT, Section 2) that seem to confirm the validity of the above method (Section~4). This method simultaneously provides a way to constrain the nebular parameters of galaxies at 2 $<$ z $<$ 7 (Section~5). Moreover, the same method could be utilized for different redshift ranges and different photometric bands. 

We assume a Chabrier initial mass function \citep[IMF, ][]{Chabrier2003}. We use WMAP7 cosmology \citep{Komatsu2011}. Finally, we assume a Solar metallicity Z$_{\odot}$=0.014 from \cite{Asplund2009}. 





\section{The SPT galaxy sample}

\citet{Reuter2020} presented the final spectroscopic redshift analysis of a flux-limited (S$_{870 \si{\um}}$ > 25 mJy) sample of galaxies from the 1.4 mm SPT survey. In this 2500 square degree survey observed at 1.4mm and 2.0mm, they identified 81 strongly lensed, dusty star-forming galaxies (DSFGs) at 1.9 $\lesssim$ z $\lesssim$ 6.9. The spectroscopic observations were conducted with the Atacama Large Millimeter/submillimeter Array (ALMA) across the 3 mm spectral window, targeting carbon monoxide line emission. From them, spectroscopic redshifts have been estimated by combining ALMA data with ancillary data. They are used in the following of this paper, and we do not estimate them when fitting the observed spectral energy distributions (SED), or when estimating the physical parameters.

With APEX/LABOCA, \cite{Strandet2016} obtained 870 $\si{\mu}$m flux densities, in the period 2010 September - 2012 November. The Herschel/SPIRE maps at 250 $\si{\mu}$m 350 $\si{\mu}$m, and 500$\si{\mu}$m were observed in two observing programs, in the period 2012 August – 2013 March. As described in \cite{Strandet2016}, the flux densities were extracted by fitting a Gaussian to the source. The peak of the Gaussian is taken as the flux density. The noise was estimated by taking the RMS in the central few arcmins of the map which is then added in quadrature to the uncertainty due to pixelation. The data themselves are from \cite{Reuter2020}.

This SPT galaxy sample contains bright IR galaxies that might include an active galactic nuclei (AGN). So a preliminary check is needed for AGN can substantially modify the IR spectral energy distribution of galaxies. However, none of the attempted observations were able to confirm the presence of an AGN in these objects, or at least a strong AGN. \cite{DeBreuck2019} derived a low AGN fraction\footnote{frac$_{AGN}$, the AGN fraction, is defined as $\frac {L_{AGN}} {L_{AGN}+L_{dust}}$} (frac$_{AGN}$ $<$ 5\%) for SPT~0418-47, and they claim that this is consistent with the AGN fractions estimated in other DSFGs from the SPT sample. In the galaxy SPT~2132-58, \cite{Bethermin2016} identified an evolved interstellar medium (ISM) with 0.5 $<$  Z/Z$_{\odot}$ $<$ 1.5, dominated by PDRs. Its CO spectral line energy distribution does not allow to reach a conclusion on the presence or not of an AGN. \cite{Gururajan2021} showed that the presence of an AGN in SPT~0103-45 is unlikely \citep[see also,][]{Ma2016, Spilker2018}. Finally, \cite{Apostolovski2019} found no signs for a signature pointing to the presence of an AGN in SPT~0346-52. 

In conclusion, for the sake of the present paper, we make the assumption that there are no major AGNs in the studied sample. 

\section{A colour-colour approach for the selection of galaxies at z $>$4}


We model the Herschel and APEX/LABOCA colours $\log_{10} \frac {PSW_{250 \si{\um}}} {PMW_{350 \si{\um}}}$ and $\log_{10} \frac {LABOCA_{870 \si{\um}}} {PLW_{500 \si{\um}}}$ with CIGALE \citep{Burgarella2005, Noll2009, Boquien2019}. With the input CIGALE parameters listed in Table~\ref{Tab.CIGALE.InputParams}, CIGALE creates 88 million models. This particular set of colours is selected to try and identify line-boosted galaxies at z~$\gtrsim$~4~-~6 as outliers in the above colour-colour diagramme.

\begin{table*}[htp]
\begin{center}
 \resizebox{0.8\linewidth}{\height}{
\begin{tabular}{|>{\centering}p{4.0cm}|>{\centering\arraybackslash}p{3.0cm}|>{\centering\arraybackslash}p{3.5cm}|>{\centering\arraybackslash}p{3.5cm}|}
 
  \hline\hline
  {\bf Parameters} & {\bf Symbol} & {\bf Fit w/o lines} & {\bf Fit w/ lines} \\
  \hline
 Target  sample    &                      &  SPT DSFGs & SPT DSFGs  \\ 
  \hline\hline
    \multicolumn{4}{c}{}\\
    \multicolumn{4}{c}{\bf Delayed SFH and recent burst}\\
  \hline
 e-folding time scale of the delayed SFH & $\tau_{main}$ [Myr] & 500 & 500 \\ 
  \hline
 Age of the main population & Age$_{main}$[Myr]  & 10 & 10 \\ 
  \hline
 Burst & f$_{burst}$  &  No burst  &  No burst  \\ 
  \hline
    \multicolumn{4}{c}{}\\
    \multicolumn{4}{c}{\bf SSP}\\
  \hline
  SSP &   & BC03 & BC03 \\ 
  \hline
  Initial mass function &  IMF & Chabrier & Chabrier \\ 
  \hline
  Metallicity     & Z &  0.02 &  0.02 \\ 
  \hline
    \multicolumn{4}{c}{}\\
    \multicolumn{4}{c}{\bf Nebular emission}\\
  \hline
  Ionization parameter &  logU  & ---  & 31 values in [-4.0, -1.0] with $\delta\log$U = 0.1 \\
  \hline
  Gas metallicity &  zgas     & --- & 0.0001, 0.001, 0.0025, 0.005, 0.007, 0.008, 0.011, 0.014, 0.016, 0.019, 0.022, 0.025, 0.03, 0.033, 0.037, 0.041, 0.046, 0.051\\
  \hline
  Electron density & ne  & ---  & 10, 100, 1000 \\
  \hline
  Line width [km/s]    &     ---    &  --- &  200 \\
  \hline
    \multicolumn{4}{c}{}\\
    \multicolumn{4}{c}{\bf Dust attenuation law (dustatt\_modified\_CF00)}\\
  \hline
  V-band attenuation in the interstellar medium&  Av\_ISM & 10 & 10 \\ 
  \hline
  Av\_ISM / (Av\_BC+Av\_ISM) &  mu &  0.44 & 0.44 \\ 
  \hline
  Power law slope of the attenuation in the ISM & slope\_ISM & -0.7 & -0.7 \\ 
  \hline
  Power law slope of the attenuation in the birth clouds & slope\_BC & -1.3 & -1.3 \\
  \hline
    \multicolumn{4}{c}{}\\
    \multicolumn{4}{c}{\bf Dust emission (casey2012)}\\
  \hline
  Temperature of the dust in K & temperature &  31 values in [30., 90.] & 31 values in [30., 90.]  \\ 
  \hline
  Emissivity index of the dust & beta & 21 values in [1.0, 3.5]  & 21 values in [1.0, 3.5]  \\ 
  \hline
  Mid-infrared powerlaw slope & alpha & 2.0 & 2.0  \\ 
  \hline
    \hline
    \multicolumn{4}{c}{}\\
    \multicolumn{4}{c}{\bf redshifting}\\
  \hline
  redshift & redshift & 81 values in [0.0, 8.0]  & 81 values in [0.0, 8.0]  \\ 
  \hline
    \multicolumn{4}{c}{}\\
    \multicolumn{4}{c}{\bf No AGN emission}\\
\hline\hline
\end{tabular}}
  \caption{CIGALE modules and input parameters used for to create the 88 million models. CF00 means \cite{Charlot2000}, BC03 means \cite{Bruzual2003}, Casey2012 means \cite{Casey2012}, and the Chabrier IMF refers to \cite{Chabrier2003}.}
  \label{Tab.CIGALE.InputParams}
\end{center}
\end{table*}

\begin{figure*}
  \centering
  \includegraphics[width=0.90\textwidth, angle=0]{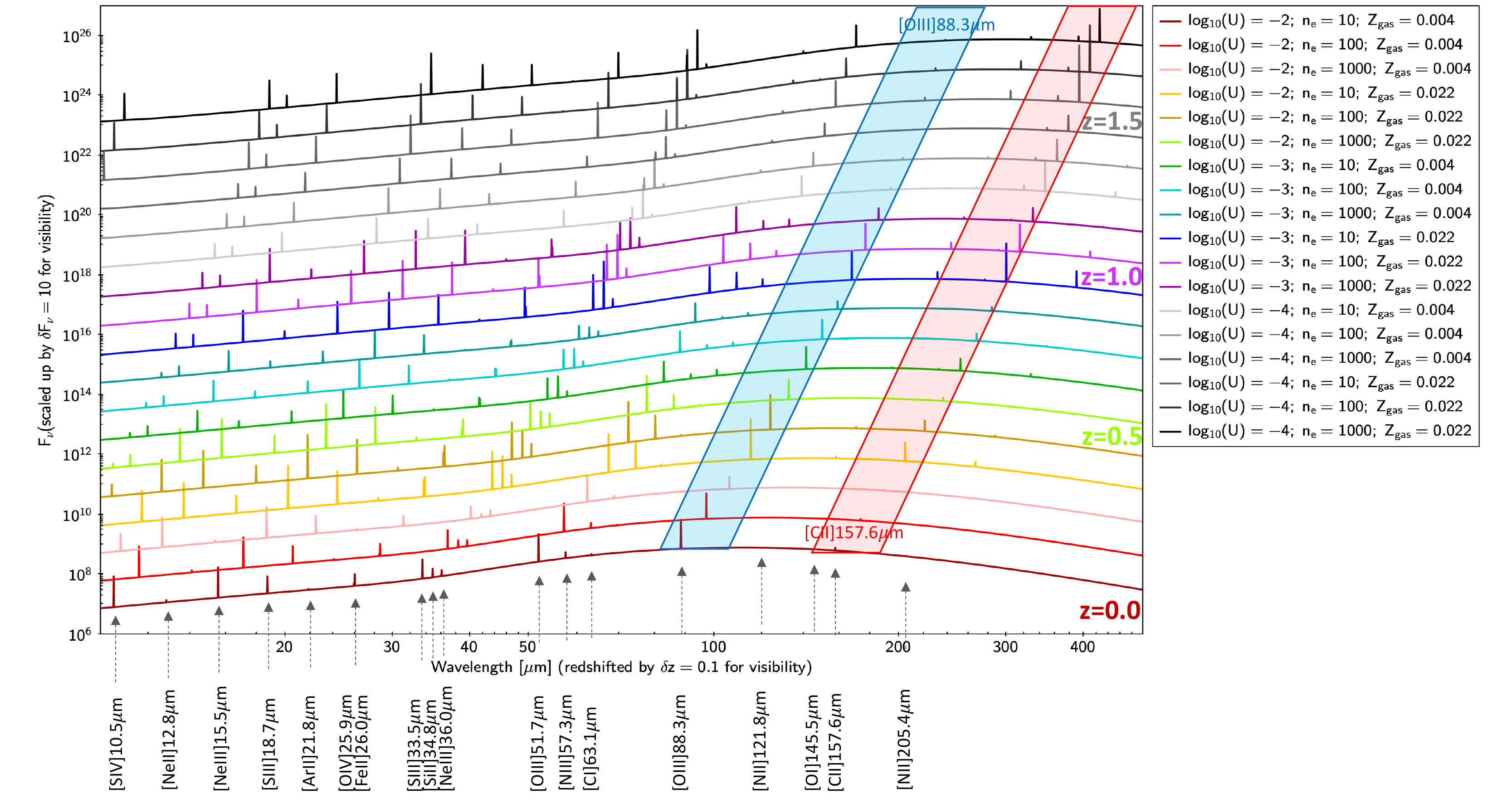}
  \caption{This sample of models created by CIGALE is an extract of the entire and much larger one used for the SED fitting ($88\times10^6$ models). Here, in addition to the nebular parameters given in the legend, we fix the SFH (delayed with $\tau_{main}$=500 Myrs, and age$_{main}$=100 Myrs), A$_V$(ISM)=0.3, T$_{dust}$=40K, the dust emissivity $\beta$=2.0, and the redshift z=0. However, to improve the visibility, we have offset the spectra by $\delta$z=0.1 on the X axis and by 1~dex on the Y axis. The bottom spectra, with $\log_{10}$ U =-2 are more representative of the IR spectra emitted by a HII-region dominated galaxy, with a strong [OIII]88.3 \si{\um} line (blue-shaded area) detected, e.g., in Lyman break galaxies in the early universe. The top spectra, with $\log_{10}$ U = -4 resemble an IR spectra emitted by a PDR dominated galaxy, with a strong [CII]157.6 \si{\um} line (red-shaded area), detected, e.g., in DSFGs in the early universe. A higher gas metallicity amplifies the strength of the metal lines.}
  \label{Fig.sample_spectra}
\end{figure*}

A grid of nebular emission lines that includes HII regions and PDRs are pre-computed with CLOUDY \citep{Ferland2017} by varying the nebular parameters $N_e$, $Z_{gas}$, and U. These emission lines are included into CIGALE when building the modelled spectra (Fig.\ref{Fig.sample_spectra}). The photo-ionizing field shape is generated with the single stellar population (SSP) model library \citep{Bruzual2003} using a constant star formation history (SFH) over 10 Myr, and accounting for a range of metallicities, ionization parameters, and number densities of hydrogen. The radiation field intensity is given by the dimensionless ionization parameter $U \equiv n_\gamma /n_H$ , where $n_\gamma$ is the number density of photons capable of ionizing hydrogen, and n$_H$ the number density of hydrogen, which equals the number density of electrons in a fully ionized medium. The limit of the effective HII region is set by an ionization fraction $\leq 10^{-3}$, while the end of the effective PDR is set by the visual attenuation $A_V \leq 10$  \citep{RolligAA07}, and as such includes part of the molecular region. The size and abundance distribution of grains, typical for the ISM of our galaxy, is used to account for the extinction in the PDR;  it includes both a graphitic and a silicate component with $R_V \equiv A_V/E(B-V) = 3.1$. The grain density scales with the hydrogen density $n_H$. The line fluxes are re-scaled with the number of Lyman continuum photons, extracted from the stellar emission of the modelled galaxies. These line models will be detailed in Theulé et al. (in prep.).

\begin{figure*}
  \centering
  \includegraphics[width=0.49\textwidth, angle=0]{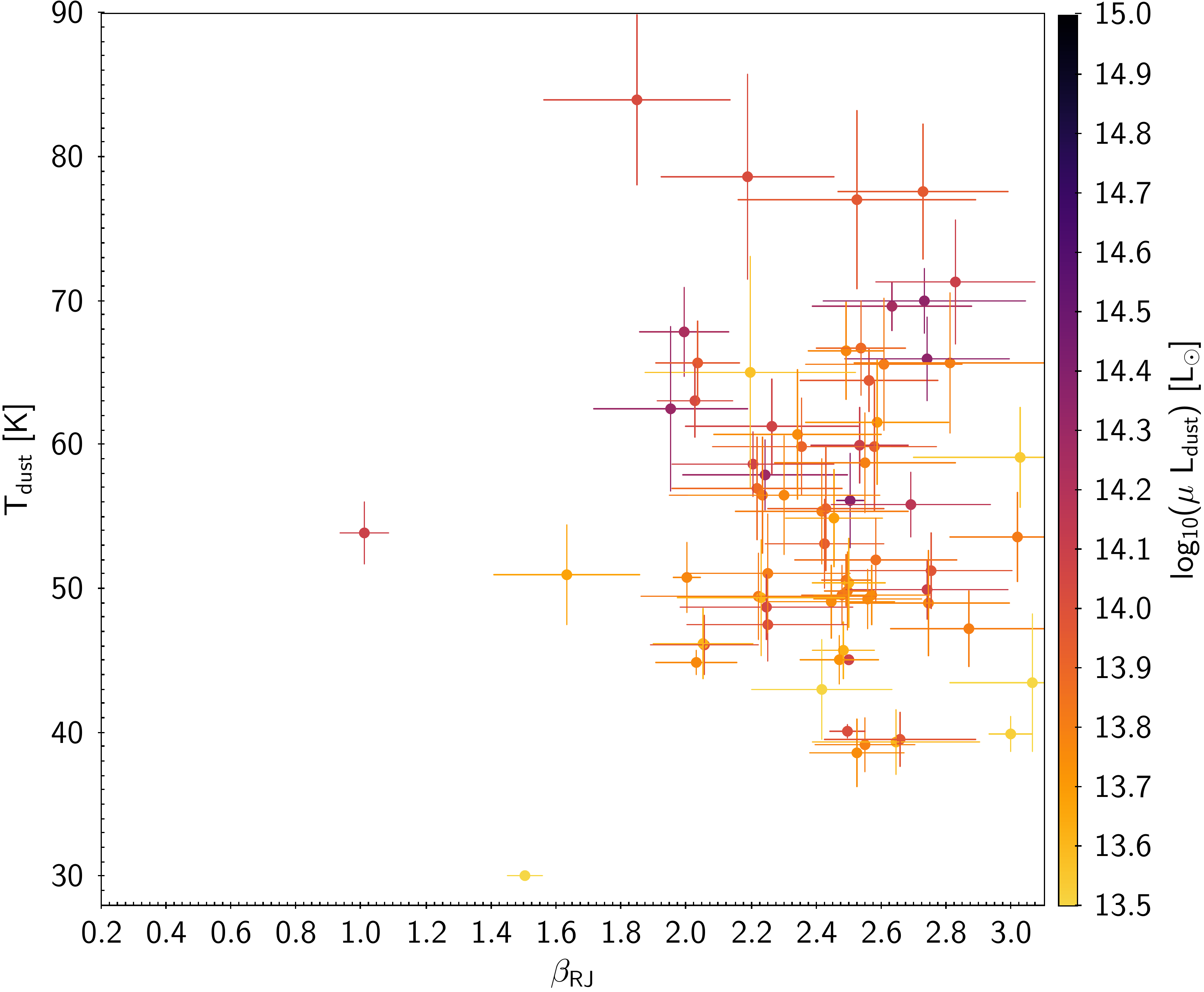}\\
  \includegraphics[width=0.49\textwidth, angle=0]{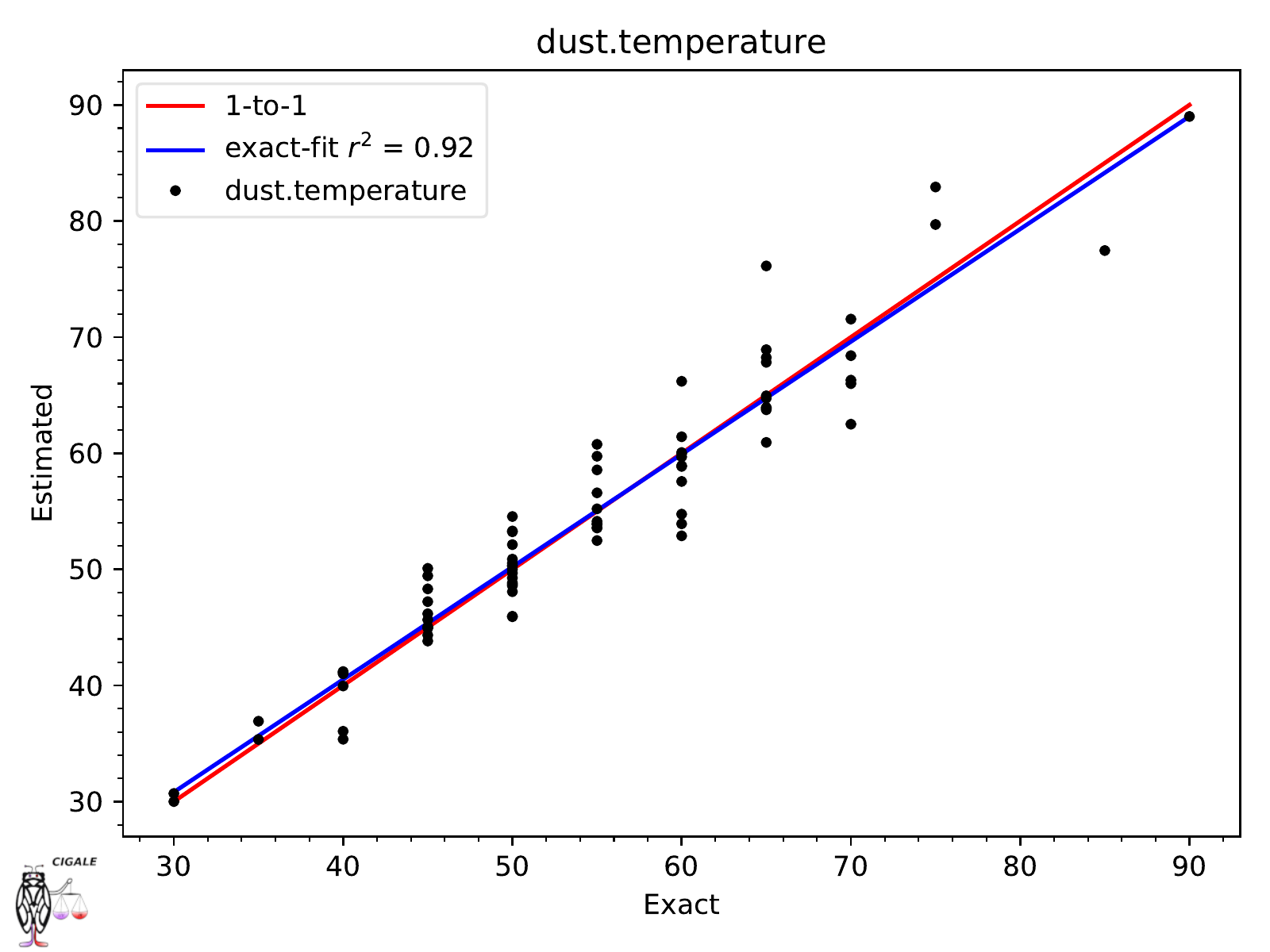}
  \includegraphics[width=0.49\textwidth, angle=0]{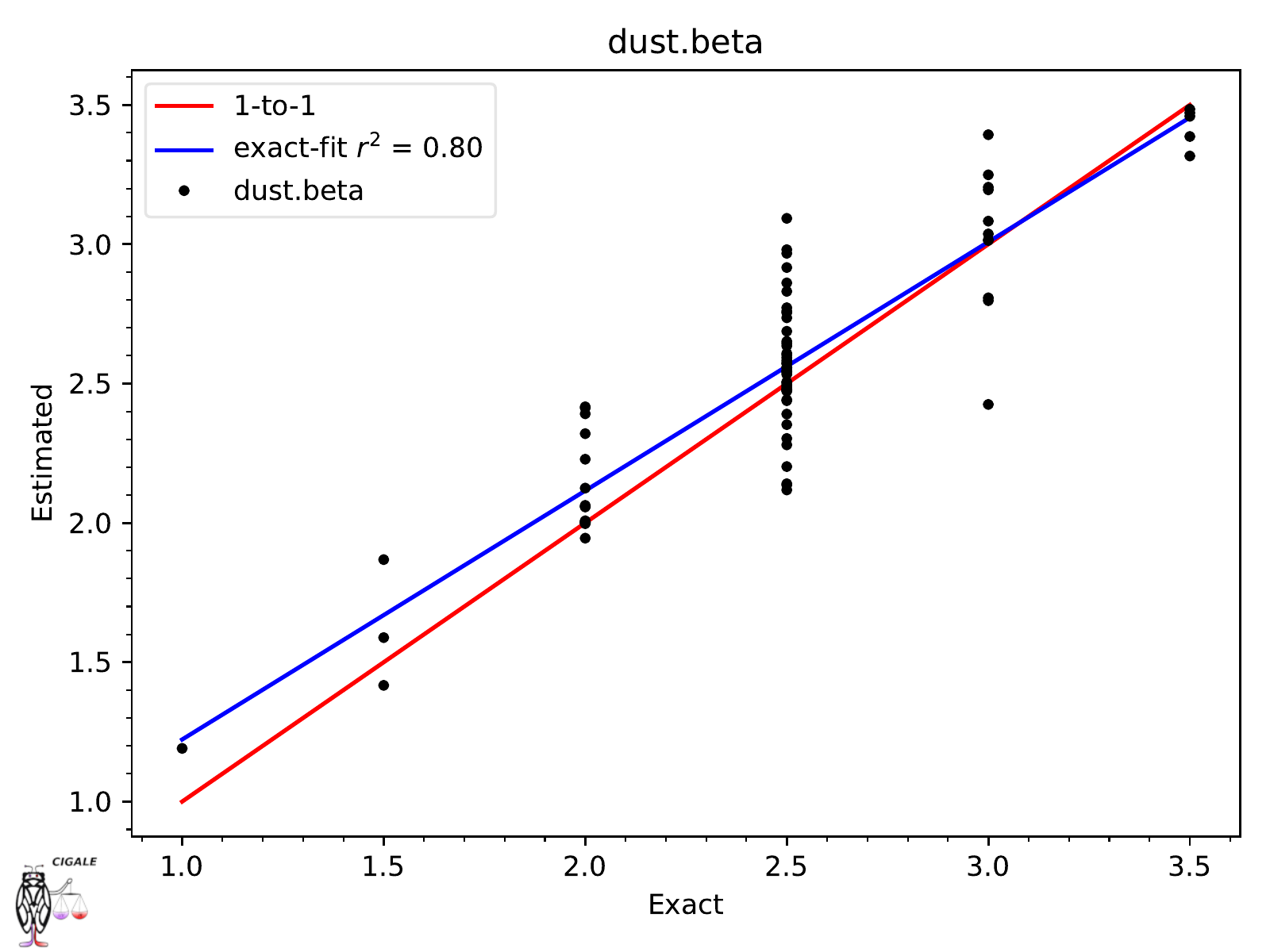}
 \caption{ The top panel shows the distribution of T$_{dust}$ and $\beta_{RL}$ for the SPT sample. In the bottom panel, we used the CIGALE mock analysis to check whether T$_{dust}$ and $\beta_{RL}$ can be correctly estimated using the available set of data for this SPT sample. In this mock analysis, we use the best models and each of the exact parameters used to build the models, for each of the objects.  We add the observed noise to the models, and refit the simulated data to re-estimate the same $_{dust}$ and $\beta_{RL}$ parameters. A good correlation between the « estimated » and the « exact » T$_{dust}$ and $\beta_{RL}$ suggests that we are able to derive them correctly.}
  \label{Fig.dust_params}
\end{figure*}

For the dust emission, we selected the modified blackbody module in CIGALE, with a power law in the mid-IR \citep{Casey2012}. In Fig.~\ref{Fig.dust_params}, we present the dust temperature and the Rayleigh-Jeans slope (T$_{dust}$ and $\beta_{RL}$) controlling the shape of the IR continuum SEDs that are derived by fitting the observed SEDs with CIGALE. T$_{dust}$ and $\beta_{RL}$ derived by fitting the data are only meant to check their measured range. They are not used in the rest of the analysis. The CIGALE mock analysis, also presented in Fig.~\ref{Fig.dust_params}, suggests that T$_{dust}$ and $\beta_{RL}$ can be well estimated, with a coefficient of correlation of r$^2=0.92$ for T$_{dust}$ and r$^2=0.80$ for $\beta_{RL}$.

\begin{figure*}[ht]
  \centering
  \includegraphics[width=0.49\textwidth,height=0.49\textwidth, angle=0]{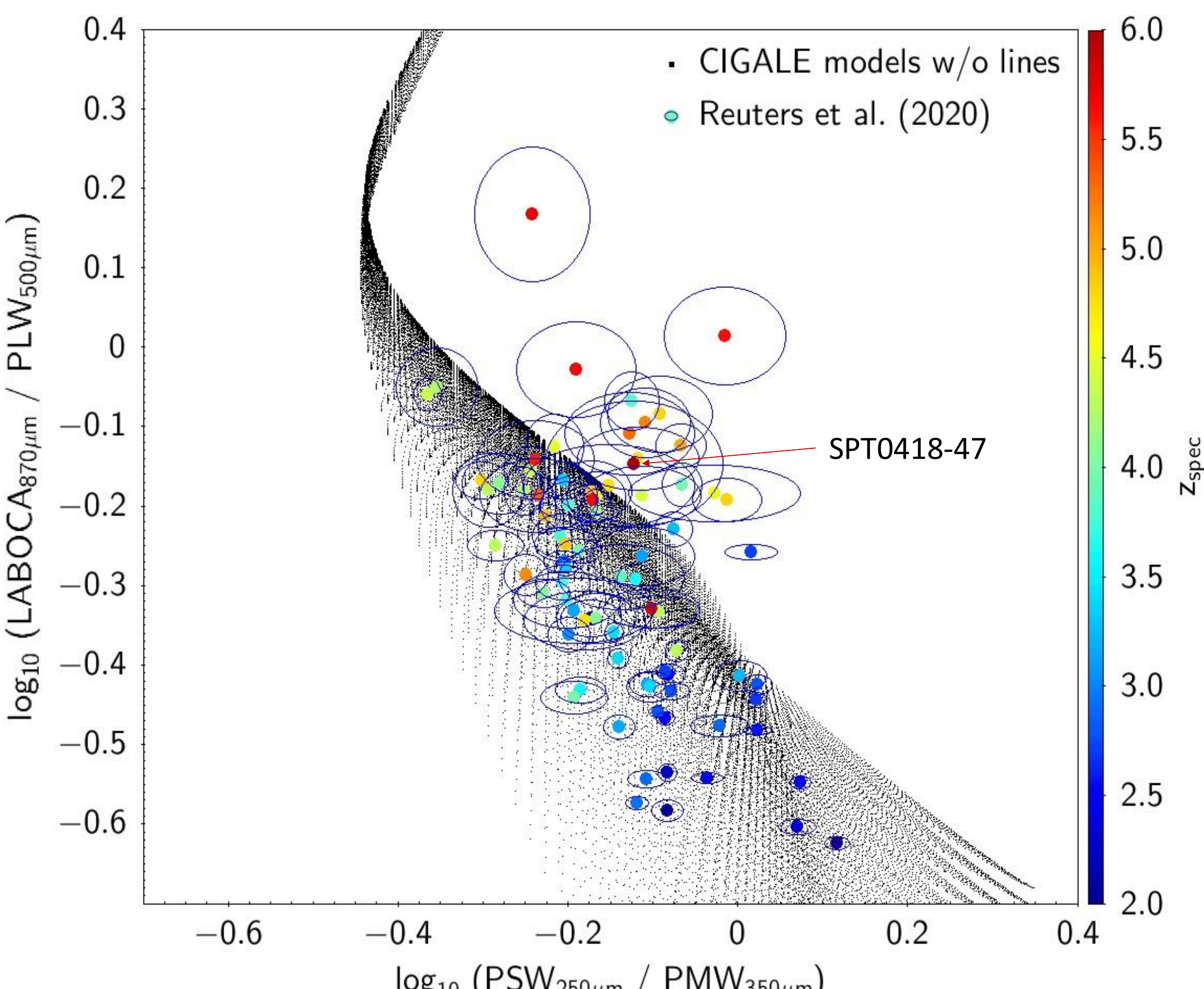}
  \includegraphics[width=0.49\textwidth,height=0.49\textwidth, angle=0]{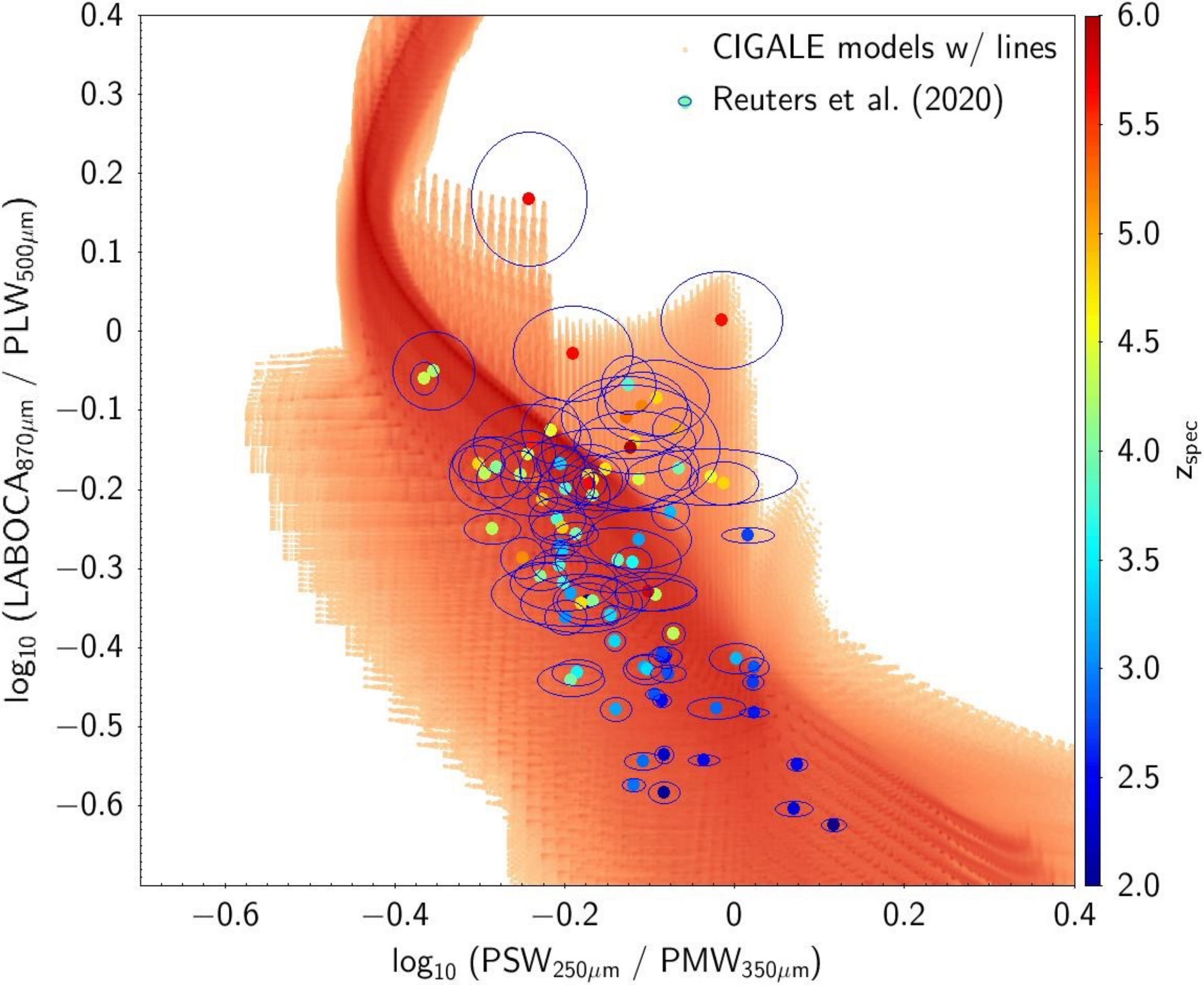}
  \caption{The models computed by CIGALE for the [$\log_{10} \frac {PSW_{250 \si{\um}}} {PMW_{350 \si{\um}}}$ versus $\log_{10} \frac {LABOCA_{870 \si{\um}}} {PLW_{500 \si{\um}}}$] colours are shown. Left: When no fine structure emission lines are added to the modelled spectra, the modelled galaxies are mainly located on a prime sequence extending from the bottom-right to the top-left. However, a clump of high redshift observed galaxies, with colours [$\log_{10} \frac {PSW_{250 \si{\um}}} {PMW_{350 \si{\um}}}$ versus $\log_{10} \frac {LABOCA_{870 \si{\um}}} {PLW_{500 \si{\um}}}$] $\approx$ [-0.1, -0.1], and with a total offset from the main sequence of the order of $\Delta_{colour} \sim 0.1 - 0.2$  cannot be reached by these models without emission lines. Right: Emission lines are added to the continuum. When large equivalent width fine structure lines (and more specifically, EW([CII]157.6 \si{\um}) $\sim$ 10 - 20 \si{\um}) are added (Appendix~\ref{Annex.line_contributions}), the CIGALE models cover this clump of high redshift galaxies. The three top-most points are only partially covered by the models. Even though the two top ones (namely SPT0243-49 and SPT0245-63) are clearly high-redshift objects, respectively at z=5.702 and z = 5.626, we assumed that the partial coverage is not sufficient and we do not use these two objects  in the rest of the paper. For the third one (SPT0348-62 at z = 5.654), the coverage by the models allows to keep it in the analysis. We also identify SPT0418-47, for which we have 5 emission lines in the considered wavelength range, and an amplification factor $\mu$ = 32.70 \citep{Reuter2020}.}
  \label{Fig.CIGALE_color_models}
\end{figure*}

In Fig.~\ref{Fig.CIGALE_color_models}, the modelled colours are compared to the observed ones in the colour-colour diagramme with and without adding emission lines to the modelled continuum. As above, we assume a mid-IR power law and a modified blackbody with a dense sampling in dust temperature T$_{dust}$, and emissivity $\beta_{RL}$ to best reproduce the rest-frame far-IR dust continuum emission. 

The main apparent trend observed in the colour-colour diagramme without emission lines, is a redshift-related sequence from the bottom-right to the top-left of the plot. This prime sequence is due to the peak of the IR dust emission, passing in the broad bands \citep[see for instance,][]{Amblard2010}. When no fine structure emission lines are added in CIGALE models, all the models are located in, or just below this prime sequence. However, we can see in  Fig.~\ref{Fig.CIGALE_color_models} that a clump of objects is located above the prime sequence, at [$\log_{10} \frac {PSW_{250 \si{\um}}} {PMW_{350 \si{\um}}}$, $\log_{10} \frac {LABOCA_{870 \si{\um}}} {PLW_{500 \si{\um}}}$] $\approx$ [-0.1, -0.1], with a total offset from the main sequence of the order of $\Delta_{colour} \sim 0.10-0.20$. This  clump contains most of the highest redshift galaxies at z $\gtrsim$ 4 - 6. They cannot be explained by changes in the dust continuum only. We quantitatively explore in Appendix~\ref{Annex.line_contributions} the possibility that emission lines are the most important piece of this puzzle. We find that galaxies exhibiting large [CII]157.6 \si{\um} equivalent width (EW) reaching EW([CII]157.6 \si{\um})  $\sim$ 10 - 20 \si{\um} could explain such outliers. This is in agreement with the estimates from \cite{Smail2011} who find that the [CII]158 \si{\um} line could provide as much as 40~\% of the 850 \si{\um} broadband flux density for z~$\sim$~4~-~6 galaxies, when L$_{[CII]} / L_{dust} >$ 1~\%.

\begin{figure*}
  \centering
   \includegraphics[width=0.9\textwidth, angle=0]{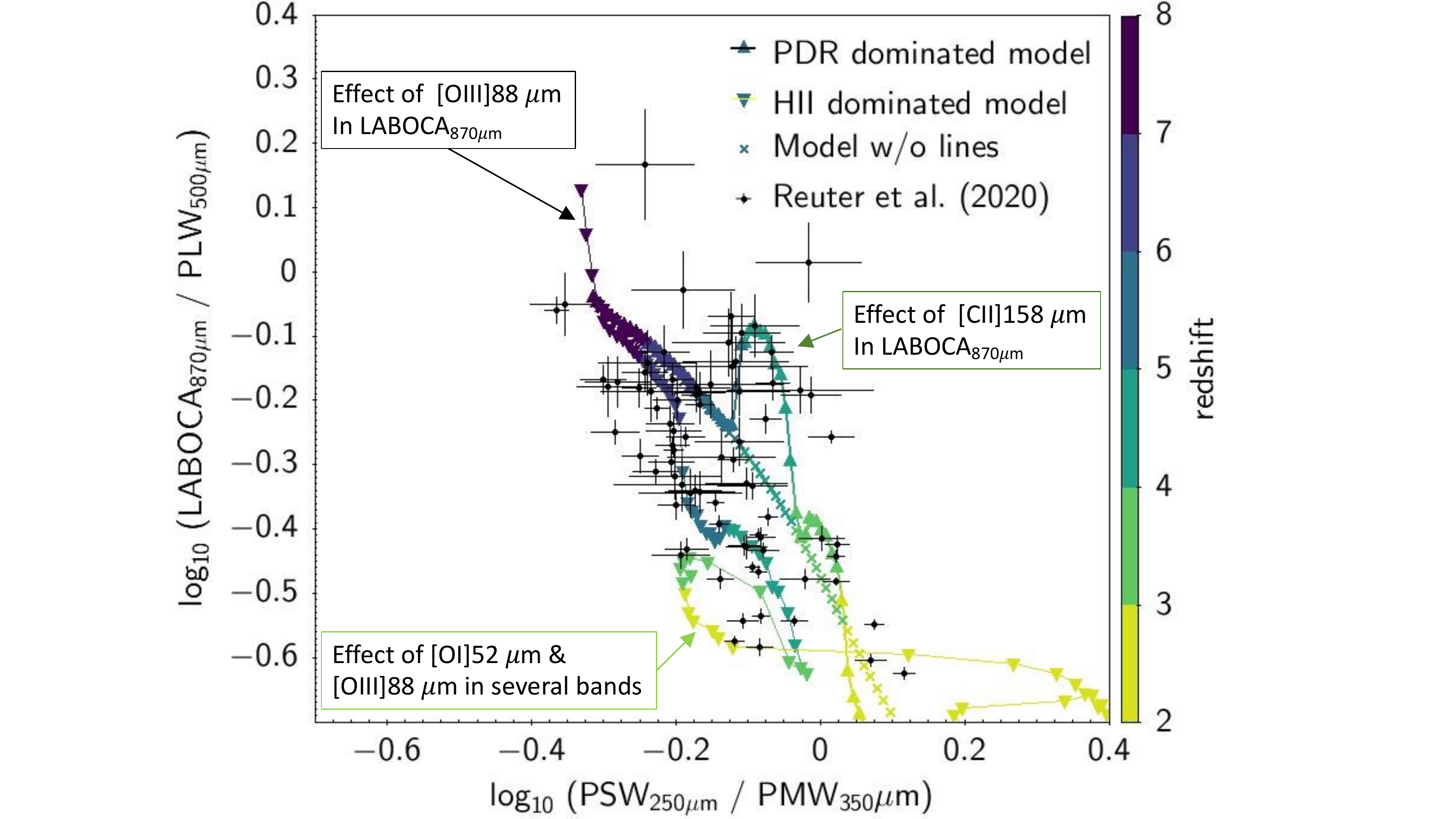}
  \caption{When compared to models without lines (crosses), at log$_{10}$ U = -4.0 (upward triangles), the strong [CII]157.6 \si{\um} line induces an upward move of $\log_{10}\frac {LABOCA_{870  \si{\um}}} {PLW_{500 \si{\um}}}$ that corresponds to the clump of galaxies at z $\sim$ 4 - 6. However, even though the move of the [OIII]51.8 \si{\um} and [OIII]88.3 \si{\um} lines of galaxies with a strong emission from HII regions at log$_{10}$ U = -2.0 (downward triangles) could induce specific colours, the effect is less clear as models with no lines can also lie in this same region of the plot (Fig.~\ref{Fig.CIGALE_color_models}). All the symbols are color-coded in redshift.}
  \label{Fig.movesin_inCCdiagramme}
\end{figure*}

Fig.~\ref{Fig.movesin_inCCdiagramme} compares the evolution in redshift of a model with emission lines, when the PDR dominates the nebular emission ($\log_{10}$ U = -4.0), and that of a model also with emission lines, but when the HII regions dominate the nebular emission ($\log_{10}$ U = -2.0). Both are also compared to the same models (that is for the same dust temperature, T$_{dust}$, and dust emissivity on the Rayleigh-Jeans side, $\beta_{RL}$)  without lines. The [CII]157.6 \si{\um} has a strong impact on the colours of PDR-dominated models. The models with large EW([CII]157.6 \si{\um}) entering into the LABOCA$_{870 \si{\um}}$ band at z $\sim$ 4 - 6 can explain the clump of galaxies above the prime sequence. At lower redshifts and for HII models, the combination of the [OIII]51.7 \si{\um}, and [OIII]88.3 \si{\um} lines (and others at lower levels) also impacts on the broad band colours, and offsets the models to below the prime sequence on the $\log_{10}\frac {LABOCA_{870  \si{\um}}} {PLW_{500 \si{\um}}}$ axis. However, some of the models without lines overlap with this region (Fig.~\ref{Fig.CIGALE_color_models}), which makes the identification of galaxies in this redshift range less conclusive for galaxies dominated by HII regions.

\section{A colour-colour approach to estimate the nebular parameters of SPT galaxies}


We saw the impact of emission lines on the colour distributions. We now try to constrain the physical parameters driving the intensity of the emission lines. For each of the SPT objects, we compute the mean of the models that lie inside ellipses delimited by the uncertainties in both colours. We stress that, for each object, we only keep the models within $\delta$z = $\pm$ 0.1 from the spectroscopic redshift. This provides us with the mean and standard deviation of the physical parameters used to compute the models, as well as those for the line and continuum fluxes. As already mentioned before, the two top objects are very partially covered by the models in Fig.~\ref{Fig.CIGALE_color_models}: SPT0243-49 and SPT0245-63 at z=5.702 and z = 5.626, respectively. They are not used hereafter.

\begin{figure*}
   \centering  
   \includegraphics[width=8cm]{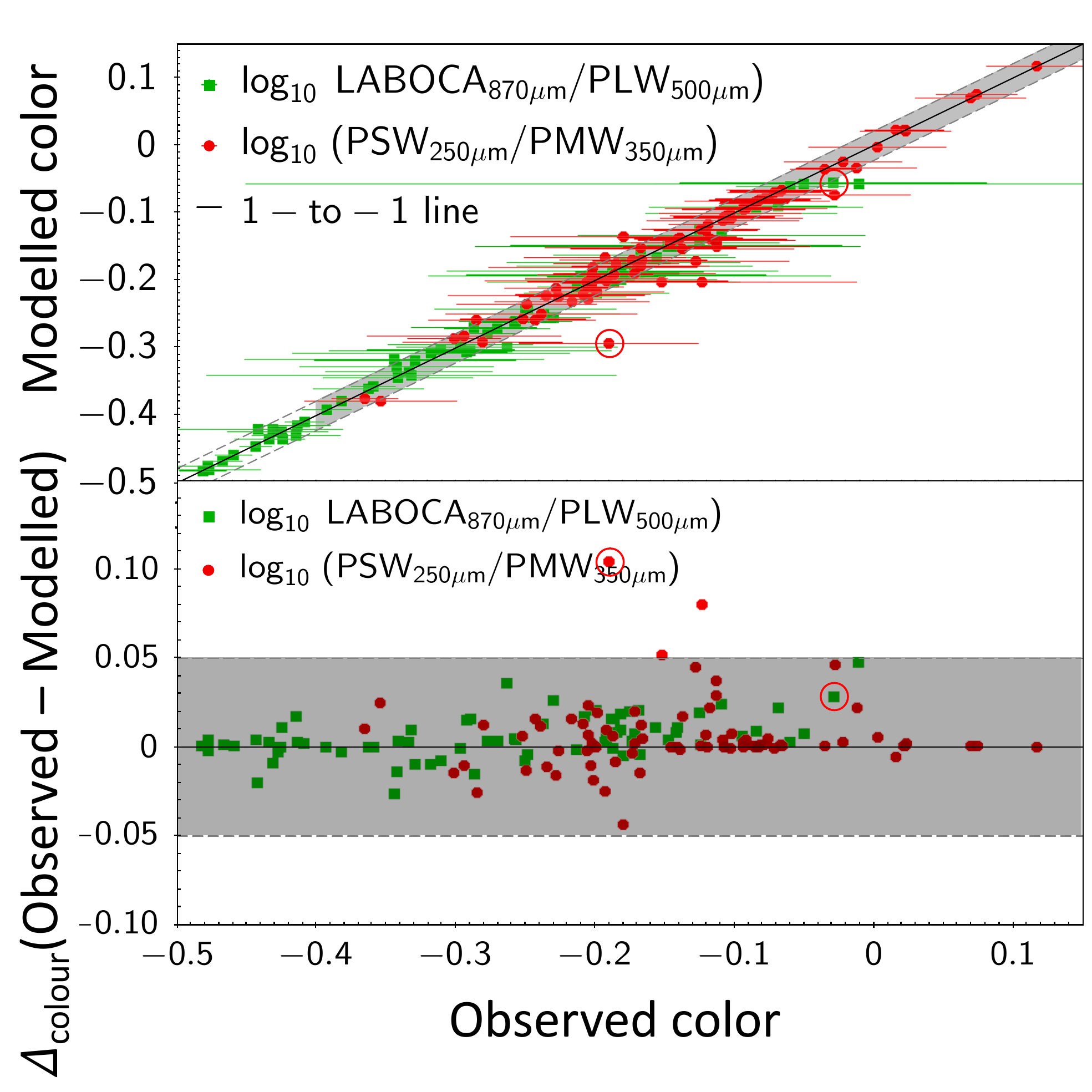}
      \caption{Top: The modelled colours (green boxes for $\log_{10}\frac {LABOCA_{870  \si{\um}}} {PLW_{500 \si{\um}}}$ and red dots for $\log_{10}\frac {PSW_{250  \si{\um}}} {PMW_{350 \si{\um}}}$) are in excellent agreement with the observed ones within the observed uncertainties. The black line shows the 1-to-1 line and the grey-shaded area presents the boundaries at  $\pm$ 5\% (y = 0.95~x and y = 1.05~x). Bottom: This conclusion is confirmed in this panel where we present the absolute differences in the logarithm of the two colors used in Fig.~\ref{Fig.CIGALE_color_models}. In both panels, the red-circled point corresponds to SPT0348-62 already identified in Fig.~\ref{Fig.CIGALE_color_models} as one of the objects only partially covered by the models.
              }
         \label{Fig.comp_colours_mod_obs}
   \end{figure*}

We compare the modelled and observed colours in Fig.~\ref{Fig.comp_colours_mod_obs}. The regressions provide correlation coefficient of 0.997 for the LABOCA\_870 \si{\um} / PLW colour and 0.957 for the PSW  / PMW colour. The median and standard deviations of the modelled-to-observed colour differences ($\Delta[\log_{10}(Colour_{Modelled}) -  \log_{10}(Colour_{Observed})]$ =  0.001 $\pm$ 0.036 for the PSW / PMW color and 0.003 $\pm$ 0.011 for the LABOCA\_870 \si{\um} / PLW colour.

\begin{figure*}
   \centering  
   \includegraphics[width=15cm]{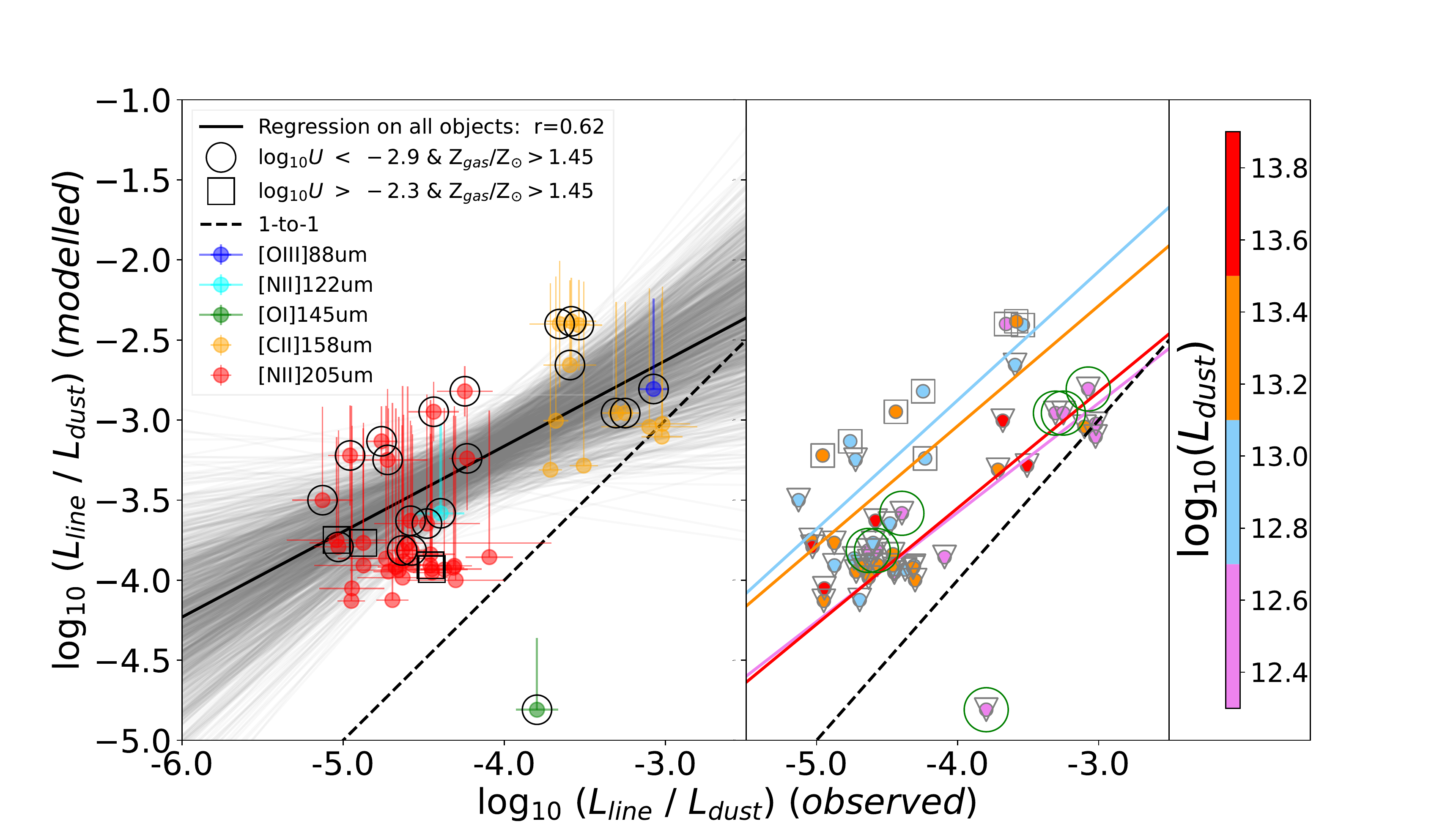}
      \caption{Left: Comparison of the observations and models for the spectroscopic sample of SPT galaxies. Most data are from [CII]157.6 \si{\um} \citep{Lagache2018} and [NII]205.2 \si{\um} \citep{Cunningham2020}, except for SPT 0418-47 \citep{DeBreuck2019} for which we have several lines. Generally speaking, the brightest [CII]157.6 \si{\um} and  [OIII]88.3 \si{\um} lines are better modelled than the [NII]122. \si{\um} and [NII]205.2 \si{\um}. The [OI]145. \si{\um} is an outlier in this frame. It could be because the former lines are stronger than the latter, or because of physical differences between them. Right: the symbols are color-coded in intrinsic dust luminosity. The fits suggest that the objects with $\log_{10}$ L$_{dust}$ $<$ 12.7, and those with $\log_{10}$ L$_{dust}$ $>$13.5 are found closer to the 1-to-1 line, or vice-versa that the line predictions for those with 12.7 $<$ $\log_{10}$ L$_{dust}$ $<$ 13.5 are worse. The objects with a large green circle corresponds to the sole object with different lines: SPT0418-47 from \cite{DeBreuck2019} (see left panel to identify the lines).
              }
         \label{Fig.comp_lines_in_out}
   \end{figure*}

In order to check our results, we compare the observed emission line fluxes of the SPT sample to our modelled ones in Fig.~\ref{Fig.comp_lines_in_out}. However, this comparison is limited because only a small sample of SPT galaxies has been spectroscopically observed, and most of them only with one line.

For a sub-sample of the SPT galaxies, \citet{Lagache2018} agree that [CII]157.6 \si{\um} at high redshift mainly originates from the PDR. For the objects in common with the present sample, their [CII]157.6 \si{\um} luminosities are compared to our derived [CII]157.6 \si{\um} luminosities. \cite{Cunningham2020} showed that 57\% of the SPT sample presents an L$_{[CII]157.6 \si{um}}$ / L$_{[NII]205.2 \si{\um}}$ luminosity ratio (or lower limit) in agreement with those expected from PDR (or shock regions). However, they suggest that a sub-sample ($\sim$27\%) of the 3 $<$ z $<$ 6 SPT galaxies would be consistent (within uncertainties) with a hybrid regime between the model predictions of PDR emission and H II regions. For the objects in common with the present sample, their [NII]205.2 \si{\um} luminosities are compared to our derived [NII]205.2 \si{\um} luminosities.

Statistical tests are performed with the Python LINMIX library (Fig.~\ref{Fig.comp_lines_in_out}). This LINMIX method presents the advantage of using a hierarchical Bayesian approach for the linear regression, that take errors in both X and Y into account \citep{Kelly2007}. The tests are significant (see Fig.~\ref{Fig.comp_lines_in_out}), given the number of points, and the linear correlation coefficient between the observed and modelled fluxes: r=0.62. 
We check whether the different locations of the lines, and most notably [CII]158 \si{\um} and [NII]205 \si{\um} for the present sample, are due to the physical mechanisms producing these two lines. However, we could not pin point any differences in the critical densities or ionisation potentials that could explain the differences in the figure \citep[see e.g. Fig.~2 in][]{Spinoglio2015}. We tentatively conclude that the main reason for the increased distance to the 1-to-1 line is very likely the difference in line intensity for [CII]158 \si{\um} and [NII]205 \si{\um}. In other words, this method is certainly more sensitive to strong lines (CII and OIII) than to faint lines (NII). The fact that there is about one order of magnitude offset between the observed and modelled line fluxes, for the weak lines, is a limitation for the method presented in this paper because the nebular parameters are based on line ratios. We note, however, another horizontal structure that does not seem random. The redshift does not explain this structure. The galaxies with an intrinsic (i.e. corrected for the amplification using \cite{Reuter2020}) dust luminosity ($\log_{10}$ L$_{dust}$) in the range 12.7~$<\log_{10}$ L$_{dust}$ $<$~13.5 are generally found at larger distances, above the 1-to-1 line. At the contrary, galaxies with $\log$ L$_{dust}$ $<$ 12.7 and galaxies with log $\log_{10}$ L$_{dust}$ $>$ 13.5 are significantly closer to this 1-to-1 line. No clear physical origin is identified for this differential effect, though. It could be related to the evolution of these very exotic high redshift objects. More data, and especially rest-frame UV (from JWST) and far-IR (from ALMA or NOEMA) morphologies are fundamental clues to decipher the structure of this diagramme.

From the present SPT sample, only for SPT0418-47 do we have several emission lines (identified with large open circles in the right panel of Fig.~\ref{Fig.comp_lines_in_out}) that allow to check how well we model the line ratios. Three of the L$_{line}$ / L$_{[OIII]88.3 \si{\um}}$ modelled ratios are in agreement, within a factor of 3 at most, to the observed ones:  L$_{[NII]122 \si{\um}}$ / L$_{[OIII]88.3 \si{\um}}$ = 0.048 $\pm$ 0.014, L$_{[CII]158 \si{\um}}$ / L$_{[OIII]88.3 \si{\um}}$ = 0.667 $\pm$ 0.091 and L$_{[NII]205 \si{\um}}$ / L$_{[OIII]88.3 \si{\um}}$ = 0.032 $\pm$ 0.004 while our estimates are 0.143, 0.619 and 0.085, respectively. The observed L$_{[OI]145 \si{\um}}$ / L$_{[OIII]88.3 \si{\um}}$ = 0.189 $\pm$ 0.050 while we find 0.011. This is a factor of almost 17 for [OI]145 \si{\um}, significantly larger than the other line ratios. However, \citet{DeBreuck2019} also had the same problem: their predicted range for this line is 0.02-1.9 $\times$ 10$^{10}$ L$_{\odot}$ while their detection amounts to 2.1 $\times$ 10$^{10}$ L$_{\odot}$, above the maximum predicted value. The conclusion, here as well, is that we need to collect more objects with these line ratios to be able to use a statistical approach and clarify the situation. For SPT0418-47, \cite{DeBreuck2019} derived a gas metallicity of 0.3 $<$ Z/Z$_{\odot}$ $<$1.3 and a ionization parameter -3.2 $<$ $\log_{10}$ U $<$ -2.0. These values are in reasonable agreement within the uncertainties with ours:  Z/Z$_{\odot}$ $\sim$ 1.47 $\pm$ 1.14 and  $\log_{10}$ U $\sim$ -3.02 $\pm$ 0.80. 

We compute the mean of the model parameters inside ellipses delimited by the observational uncertainties. The metallicities are in the range 0.5 $\leq$ Z/Z$_\odot$ $\leq$ 2.5 and the ionization parameters cover from -4.0 $\leq$ $\log_{10}$ U $\leq$ -1.5. (Fig.~\ref{Fig.logU_Z}). Very high metallicity objects are rare, but observed in the local universe \citep[][]{Gallazzi2005, Peeples2008, Maiolino2019} with metallicities that extend to $>$ 3 Z$_{\odot}$. Such high values for the ionization parameter were also measured in the overlap region of the Antennae galaxies by, e.g., \citet{Snijders2007, Kewley2019}. \cite{Yeh2012} found that radiation pressure confinement sets an upper limit to $\log_{10}$ U = -1 in individual regions. However, when observing unresolved starbursts the mean values are of the order of $\log_{10}$ U = -2.3, due to the variety of regions inside a galaxy. As noted earlier in this paper, the two top most objects in Fig.~\ref{Fig.CIGALE_color_models} either are very extreme cases among the diversity of galaxies, or the observed colour uncertainties are underestimated. We do not keep them the the analysis. However, they probably deserve specific studies.



 ------------------------%
\section{A structure in the $\log_{10}$ U vs. $\log_{10}$ Z/Z$_{\odot}$ diagram}

\begin{figure}
   \centering
   \includegraphics[width=9cm]{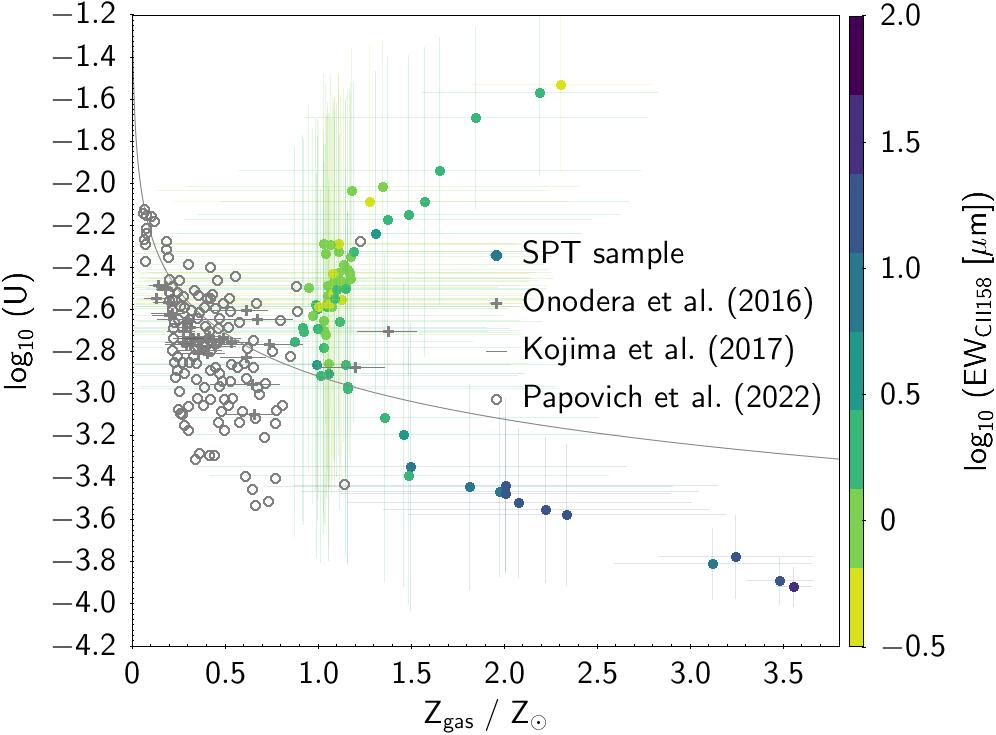}
   \includegraphics[width=9cm]{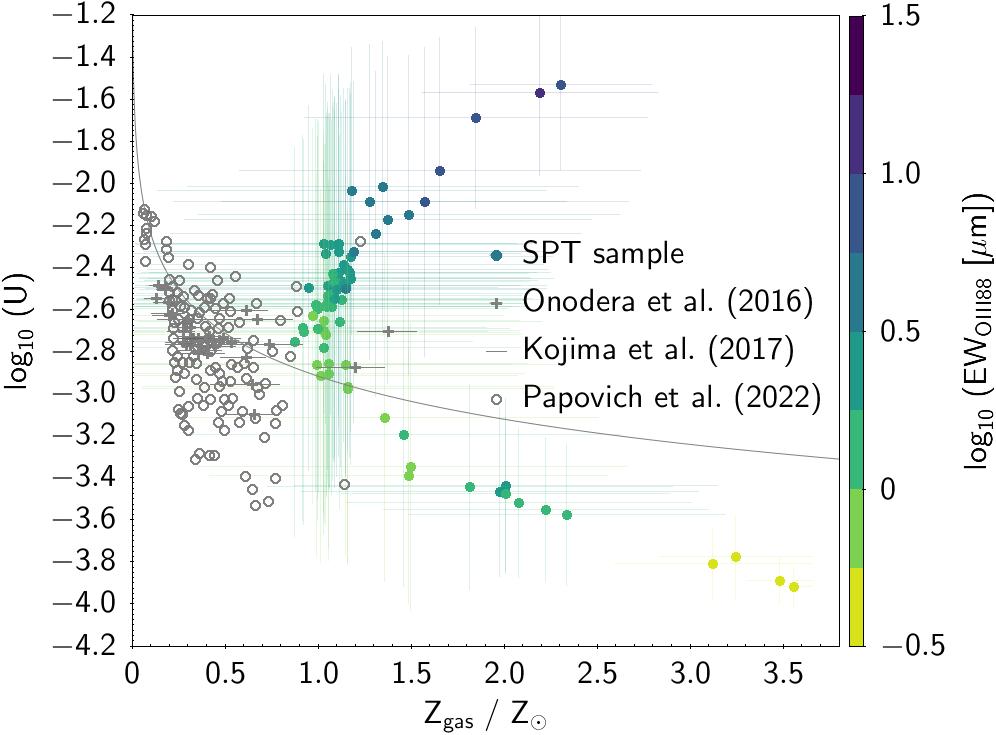}
   \includegraphics[width=9cm]{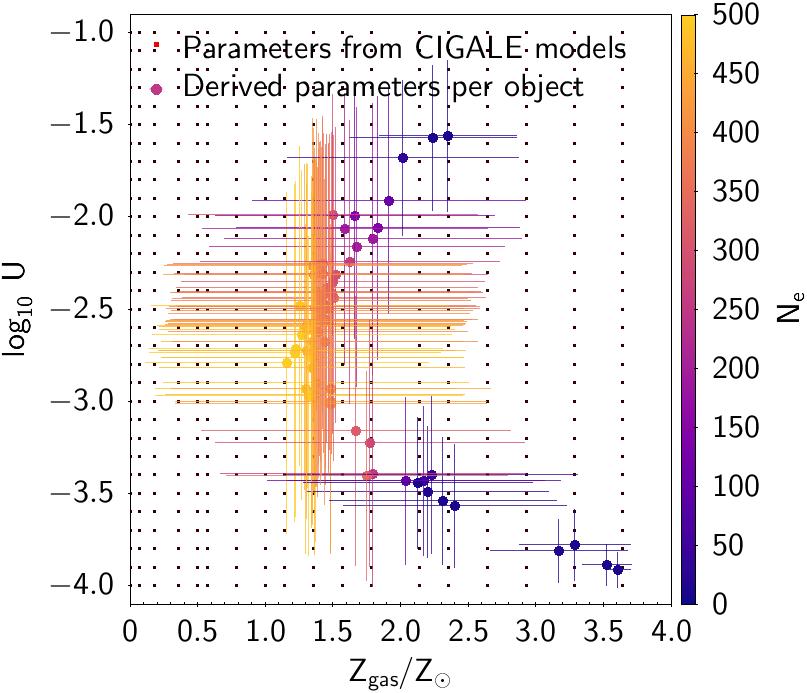}
      \caption{The sequence where we find LBGs at z $\sim$ 3.3 sample \citep{Onodera2016} and the fit (dashed line) to the sequence from \cite{Kojima2017} are shown in grey. Two different branches are identified for which the [CII]157.6 \si{\um} (centre) and [OIII]88.3 \si{\um} (top) lines are strong. The top branch contains galaxies that present a [C II] deficit and have a stronger ionization parameter, $\log_{10}$ U $\lesssim -2.5$ while the bottom  branch correspond to galaxies with a normal to extreme PDR emission. The bottom panel shows the same information color-coded with the electronic density, with the grid of models superimposed that shows that we should be able to derive metallicities and ionization parameters in between the two branches.
              }
         \label{Fig.logU_Z}
   \end{figure}

From the parameters derived in the previous section, we build the $\log_{10}$ U vs. Z/Z$_{\odot}$ diagramme (Fig.~\ref{Fig.logU_Z}), where we identify two sequences: the top one with a strong L$_{[OIII]88.3 \si{\um}}$ and the bottom one with a strong L$_{[CII]157.6 \si{\um}}$ emission. Even though the models are created on a dense regular grid, with a flat probability (bottom panel of Fig.~\ref{Fig.logU_Z}), we stress that the metallicities and ionization parameters derived from the method presented in this paper cannot be precise enough to define the two sequences as clearly as they appear in Fig.~\ref{Fig.logU_Z}. The well-defined sequences might be due to the fact that the mean values are estimated from wide probability distribution functions, as confirmed by the uncertainties shown in Fig.~\ref{Fig.logU_Z}. The mean values of these wide distributions regularly evolve in the plot, that could suggest the impression of well-defined sequences.


The bottom sequence presents $\log_{10}$ U values that are similar to PDR-dominated galaxies while they are more similar to HII region-dominated galaxies for the top sequence. This type of objects would be galaxies that present a so-called [C II] deficit. These  [C II] deficit galaxies show a ratio L$_{[OIII]88.3 \si{\um}}$ / L$_{[CII]157.6 \si{\um}}$ $\approx$ 3 -- 20, that is about 10 times higher than z~$\sim$~0 galaxies. \cite{Harikane2020} identified nine z = 6-9 galaxies whose observed properties are in agreement with being such [C~II] deficit galaxies. Numerous explanations have been proposed: differences in C and O abundance ratios, observational biases, and differences in ISM properties. \cite{Carniani2020} suggest that a surface brightness dimming of the extended [C II] emission would be responsible for the [C II] deficit. \cite{Harikane2020} explain these high L$_{[OIII]88.3 \si{\um}}$ / L$_{[CII]157.6 \si{\um}}$ ratios by high ionization parameters or low PDR covering fractions, both of which are consistent with their [N II] observations. This scenario could be reproduced by a density-bounded nebula with a PDR deficit. In radiation-hydrodynamics simulations, 
\cite{Abel2009} concluded that the effects of high ratios of impinging ionizing radiation density to particle density (i.e. again high ionization parameters) can reproduce the observational characteristics of ultra luminous IR galaxies (ULIRG). When U increases, the fraction of UV photons absorbed by dust increases, and fewer photons are available to photoionize and heat the gas. This leads to a dust-bounded nebula that can explain the observed [CII] deficit \cite[see also ][for a slightly more complex but consistent explanation]{Fischer2014}.  

\section{Conclusions}

We perform an analysis of the SEDs of SPT galaxies with spectroscopic redshifts. We built more than $88\times10^6$ models with CIGALE and compare the observed objects to the models in a [$\log_{10}$ (PSW$_{250 \si{\um}}$ / PMW$_{350 \si{\um}}$) versus $\log_{10}$(LABOCA$_{870 \si{\um}}$ / PLW$_{500 \si{\um}}$)] colour-colour diagram. This set of colours is selected to identify galaxies at z $\gtrsim$ 4 - 6 from the influence of the [CII]158 \si{\um} fine structure lines on broad bands. This method also allows to roughly estimate the nebular parameters for this SPT sample. 
 
From this analysis, we find the following results.

\begin{itemize}{}
      \item The position of the SPT z $\sim$ 4 - 6 galaxies in the colour-colour diagramme, can only be explained when adding the contribution of fine structure far-IR emission lines to the dust continuum. 
      \item By averaging the models and their associated physical parameters in ellipses delimited by the observed uncertainties in both colours, we can estimate the flux of the fine structure far-IR emission lines, the gas metallicity (Z$_{gas}$), and the ionization parameter ($\log_{10}$ U). We find that all the SPT galaxies have high gas metallicity 0.6 < Z$_{gas}$ < 2.5 and they cover a wide range of ionization parameter in the range $\log_{10}$ U = -4.0 to -1.5. However, we add a word of caution because the faintest lines are over estimated. Thus, line ratios involving faint lines could bias the estimation of nebular parameters.
      \item In the $\log_{10}$ U versus Z$_{gas}$ diagram, we identify two branches with high [OIII]88 \si{\um} / [CII]158 \si{\um} ratios for the top branch and low [OIII]88 \si{\um} / [CII]158 \si{\um} ratios for the bottom branch. The top branch, with high $\log_{10}$ U $\lesssim$ -2.5, presents a deficit in L$_{CII}$/L$_{dust}$ with respect to the bulk of the galaxy sample, while the bottom branch is the extension of a sequence that continues to LBGs at lower metallicities. 
      \item Without emission lines,  outliers are offset from the prime sequence by $\mid \Delta (\log_{10} (PSW_{250 \si{\um}} / PMW_{350 \si{\um}}) \mid $ $\approx$ 0.05 and $\mid \Delta (\log_{10}(LABOCA_{870 \si{\um}} / PLW_{500 \si{\um}}) \mid $ $\approx$ 0.09. That is a total offset of the order of $\Delta_{total} \sim 0.10$. For SPT galaxies at z $\gtrsim$ 4 - 6, the main effect is due to [CII]158 \si{\um} with 0.6~$\lesssim$~EW([CII]157.6~\si{\um})~$\lesssim$~25.0 \si{\um}. For the most extreme cases, this line could be at the origin of almost half of the flux density at 850 \si{\um}, for galaxies at z~$\sim$~4~-~6.
      \item In order to make the most efficient use of this method, a set of medium and broad bands in the mid- and far-IR would be ideal because the effect of the lines would be stronger in narrower bands. One of the caveats of this method is the need to collect an SED as complete as possible to correctly estimate the line fluxes: the more complete the SED, the better the line fluxes can be estimated. Thus, to be efficient, the utility of the method presented here relies on large photometric samples, which are cheaper to obtain than spectroscopy. In other words, our method would benefit from large photometric surveys on large galaxy sample. Otherwise, spectroscopic observations present the advantage of providing much better estimates. With this configuration, a project like the The PRobe far-Infrared Mission for Astrophysics (PRIMA\footnote{See: https://prima.ipac.caltech.edu/ and https://agora.lam.fr/}) allows a statistical approach that would permit to understand the cosmic rise of metals up to the reionization.
\end{itemize}

\begin{acknowledgements}
The authors thank Charles M. (Matt) Bradford for a very useful discussion. DB acknowledges support from the Centre National d'Etudes Spatiales (CNES) for this effort of simulating galaxies and their properties derived from space facilities like {\it Herschel} and predict new space observations. Médéric Boquien gratefully acknowledges support by the ANID BASAL project FB210003 and from the FONDECYT regular grant 1211000. A.K.I., Y.S., and Y.F. are supported by NAOJ ALMA Scientific Research grant No. 2020-16B.
\end{acknowledgements}

\bibliographystyle{aa} 
\bibliography{IR_colors_lines_arXiv} 

\begin{thebibliography}{51}
\expandafter\ifx\csname natexlab\endcsname\relax\def\natexlab#1{#1}\fi

\bibitem[{{Abel} {et~al.}(2009){Abel}, {Dudley}, {Fischer}, {Satyapal}, \& {van
  Hoof}}]{Abel2009}
{Abel}, N.~P., {Dudley}, C., {Fischer}, J., {Satyapal}, S., \& {van Hoof},
  P.~A.~M. 2009, \apj, 701, 1147

\bibitem[{{Algera} {et~al.}(2022){Algera}, {Inami}, {Oesch}, {Sommovigo},
  {Bouwens}, {Topping}, {Schouws}, {Stefanon}, {Stark}, {Aravena}, {Barrufet},
  {da Cunha}, {Dayal}, {Endsley}, {Ferrara}, {Fudamoto}, {Gonzalez},
  {Graziani}, {Hodge}, {Hygate}, {de Looze}, {Nanayakkara}, {Schneider}, \&
  {van der Werf}}]{Algera2022}
{Algera}, H., {Inami}, H., {Oesch}, P., {et~al.} 2022, arXiv e-prints,
  arXiv:2208.08243

\bibitem[{{Amblard} {et~al.}(2010){Amblard}, {Cooray}, {Serra}, {Temi},
  {Barton}, {Negrello}, {Auld}, {Baes}, {Baldry}, {Bamford}, {Blain}, {Bock},
  {Bonfield}, {Burgarella}, {Buttiglione}, {Cameron}, {Cava}, {Clements},
  {Croom}, {Dariush}, {de Zotti}, {Driver}, {Dunlop}, {Dunne}, {Dye}, {Eales},
  {Frayer}, {Fritz}, {Gardner}, {Gonzalez-Nuevo}, {Herranz}, {Hill}, {Hopkins},
  {Hughes}, {Ibar}, {Ivison}, {Jarvis}, {Jones}, {Kelvin}, {Lagache}, {Leeuw},
  {Liske}, {Lopez-Caniego}, {Loveday}, {Maddox}, {Micha{\l}owski}, {Norberg},
  {Parkinson}, {Peacock}, {Pearson}, {Pascale}, {Pohlen}, {Popescu},
  {Prescott}, {Robotham}, {Rigby}, {Rodighiero}, {Samui}, {Sansom}, {Scott},
  {Serjeant}, {Sharp}, {Sibthorpe}, {Smith}, {Thompson}, {Tuffs}, {Valtchanov},
  {van Kampen}, {van der Werf}, {Verma}, {Vieira}, \& {Vlahakis}}]{Amblard2010}
{Amblard}, A., {Cooray}, A., {Serra}, P., {et~al.} 2010, \aap, 518, L9

\bibitem[{{Anders} \& {Fritze-v. Alvensleben}(2003)}]{Anders2003}
{Anders}, P. \& {Fritze-v. Alvensleben}, U. 2003, \aap, 401, 1063

\bibitem[{{Apostolovski} {et~al.}(2019){Apostolovski}, {Aravena}, {Anguita},
  {Spilker}, {Wei{\ss}}, {B{\'e}thermin}, {Chapman}, {Chen}, {Cunningham}, {De
  Breuck}, {Dong}, {Hayward}, {Hezaveh}, {Jarugula}, {Litke}, {Ma}, {Marrone},
  {Narayanan}, {Reuter}, {Rotermund}, \& {Vieira}}]{Apostolovski2019}
{Apostolovski}, Y., {Aravena}, M., {Anguita}, T., {et~al.} 2019, \aap, 628, A23

\bibitem[{{Asplund} {et~al.}(2009){Asplund}, {Grevesse}, {Sauval}, \&
  {Scott}}]{Asplund2009}
{Asplund}, M., {Grevesse}, N., {Sauval}, A.~J., \& {Scott}, P. 2009, \araa, 47,
  481

\bibitem[{{B{\'e}thermin} {et~al.}(2016){B{\'e}thermin}, {De Breuck},
  {Gullberg}, {Aravena}, {Bothwell}, {Chapman}, {Gonzalez}, {Greve}, {Litke},
  {Ma}, {Malkan}, {Marrone}, {Murphy}, {Spilker}, {Stark}, {Strandet},
  {Vieira}, {Wei{\ss}}, \& {Welikala}}]{Bethermin2016}
{B{\'e}thermin}, M., {De Breuck}, C., {Gullberg}, B., {et~al.} 2016, \aap, 586,
  L7

\bibitem[{{Boquien} {et~al.}(2019){Boquien}, {Burgarella}, {Roehlly}, {Buat},
  {Ciesla}, {Corre}, {Inoue}, \& {Salas}}]{Boquien2019}
{Boquien}, M., {Burgarella}, D., {Roehlly}, Y., {et~al.} 2019, \aap, 622, A103

\bibitem[{{Bruzual} \& {Charlot}(2003)}]{Bruzual2003}
{Bruzual}, G. \& {Charlot}, S. 2003, \mnras, 344, 1000

\bibitem[{{Burgarella} {et~al.}(2005){Burgarella}, {Buat}, \&
  {Iglesias-P{\'a}ramo}}]{Burgarella2005}
{Burgarella}, D., {Buat}, V., \& {Iglesias-P{\'a}ramo}, J. 2005, \mnras, 360,
  1413

\bibitem[{{Carniani} {et~al.}(2020){Carniani}, {Ferrara}, {Maiolino},
  {Castellano}, {Gallerani}, {Fontana}, {Kohandel}, {Lupi}, {Pallottini},
  {Pentericci}, {Vallini}, \& {Vanzella}}]{Carniani2020}
{Carniani}, S., {Ferrara}, A., {Maiolino}, R., {et~al.} 2020, \mnras, 499, 5136

\bibitem[{{Casey}(2012)}]{Casey2012}
{Casey}, C.~M. 2012, \mnras, 425, 3094

\bibitem[{{Chabrier}(2003)}]{Chabrier2003}
{Chabrier}, G. 2003, \pasp, 115, 763

\bibitem[{{Charlot} \& {Fall}(2000)}]{Charlot2000}
{Charlot}, S. \& {Fall}, S.~M. 2000, \apj, 539, 718

\bibitem[{{Cunningham} {et~al.}(2020){Cunningham}, {Chapman}, {Aravena}, {De
  Breuck}, {B{\'e}thermin}, {Chen}, {Dong}, {Gonzalez}, {Greve}, {Litke}, {Ma},
  {Malkan}, {Marrone}, {Miller}, {Phadke}, {Reuter}, {Rotermund}, {Spilker},
  {Stark}, {Strandet}, {Vieira}, \& {Wei{\ss}}}]{Cunningham2020}
{Cunningham}, D.~J.~M., {Chapman}, S.~C., {Aravena}, M., {et~al.} 2020, \mnras,
  494, 4090

\bibitem[{{de Barros} {et~al.}(2013){de Barros}, {Nayyeri}, {Reddy}, \&
  {Mobasher}}]{deBarros2013}
{de Barros}, S., {Nayyeri}, H., {Reddy}, N., \& {Mobasher}, B. 2013, in
  SF2A-2013: Proceedings of the Annual meeting of the French Society of
  Astronomy and Astrophysics, ed. L.~{Cambresy}, F.~{Martins}, E.~{Nuss}, \&
  A.~{Palacios}, 531--535

\bibitem[{{De Breuck} {et~al.}(2019){De Breuck}, {Wei{\ss}}, {B{\'e}thermin},
  {Cunningham}, {Apostolovski}, {Aravena}, {Archipley}, {Chapman}, {Chen},
  {Fu}, {Jarugula}, {Malkan}, {Mangian}, {Phadke}, {Reuter}, {Stacey},
  {Strandet}, {Vieira}, \& {Vishwas}}]{DeBreuck2019}
{De Breuck}, C., {Wei{\ss}}, A., {B{\'e}thermin}, M., {et~al.} 2019, \aap, 631,
  A167

\bibitem[{{D{\'\i}az-Santos} {et~al.}(2017){D{\'\i}az-Santos}, {Armus},
  {Charmandaris}, {Lu}, {Stierwalt}, {Stacey}, {Malhotra}, {van der Werf},
  {Howell}, {Privon}, {Mazzarella}, {Goldsmith}, {Murphy}, {Barcos-Mu{\~n}oz},
  {Linden}, {Inami}, {Larson}, {Evans}, {Appleton}, {Iwasawa}, {Lord},
  {Sanders}, \& {Surace}}]{Diaz-Santos2017}
{D{\'\i}az-Santos}, T., {Armus}, L., {Charmandaris}, V., {et~al.} 2017, \apj,
  846, 32

\bibitem[{{D{\'\i}az-Santos} {et~al.}(2013){D{\'\i}az-Santos}, {Armus},
  {Charmandaris}, {Stierwalt}, {Murphy}, {Haan}, {Inami}, {Malhotra},
  {Meijerink}, {Stacey}, {Petric}, {Evans}, {Veilleux}, {van der Werf}, {Lord},
  {Lu}, {Howell}, {Appleton}, {Mazzarella}, {Surace}, {Xu}, {Schulz},
  {Sanders}, {Bridge}, {Chan}, {Frayer}, {Iwasawa}, {Melbourne}, \&
  {Sturm}}]{Diaz-Santos2013}
{D{\'\i}az-Santos}, T., {Armus}, L., {Charmandaris}, V., {et~al.} 2013, \apj,
  774, 68

\bibitem[{{Ferland} {et~al.}(2017){Ferland}, {Chatzikos}, {Guzm{\'a}n},
  {Lykins}, {van Hoof}, {Williams}, {Abel}, {Badnell}, {Keenan}, {Porter}, \&
  {Stancil}}]{Ferland2017}
{Ferland}, G.~J., {Chatzikos}, M., {Guzm{\'a}n}, F., {et~al.} 2017, \rmxaa, 53,
  385

\bibitem[{{Finkelstein} {et~al.}(2022){Finkelstein}, {Bagley}, {Arrabal Haro},
  {Dickinson}, {Ferguson}, {Kartaltepe}, {Papovich}, {Burgarella}, {Kocevski},
  {Huertas-Company}, {Iyer}, {Larson}, {P{\'e}rez-Gonz{\'a}lez}, {Rose},
  {Tacchella}, {Wilkins}, {Chworowsky}, {Medrano}, {Morales}, {Somerville},
  {Yung}, {Fontana}, {Giavalisco}, {Grazian}, {Grogin}, {Kewley}, {Koekemoer},
  {Kirkpatrick}, {Kurczynski}, {Lotz}, {Pentericci}, {Pirzkal}, {Ravindranath},
  {Ryan}, {Trump}, {Yang}, {Almaini}, {Amor{\'\i}n}, {Annunziatella},
  {Backhaus}, {Barro}, {Behroozi}, {Bell}, {Bhatawdekar}, {Bisigello}, {Bromm},
  {Buat}, {Buitrago}, {Calabr{\'o}}, {Casey}, {Castellano}, {Ch{\'a}vez Ortiz},
  {Ciesla}, {Cleri}, {Cohen}, {Cole}, {Cooke}, {Cooper}, {Cooray}, {Costantin},
  {Cox}, {Croton}, {Daddi}, {Dav{\'e}}, {de la Vega}, {Dekel}, {Elbaz},
  {Estrada-Carpenter}, {Faber}, {Fern{\'a}ndez}, {Finkelstein}, {Freundlich},
  {Fujimoto}, {Garc{\'\i}a-Argum{\'a}nez}, {Gardner}, {Gawiser},
  {G{\'o}mez-Guijarro}, {Guo}, {Hamilton}, {Hathi}, {Holwerda}, {Hirschmann},
  {Hutchison}, {Jaskot}, {Jha}, {Jogee}, {Juneau}, {Jung}, {Kassin}, {Le Bail},
  {Leung}, {Lucas}, {Magnelli}, {Mantha}, {Matharu}, {McGrath}, {McIntosh},
  {Merlin}, {Mobasher}, {Newman}, {Nicholls}, {Pandya}, {Rafelski}, {Ronayne},
  {Santini}, {Seill{\'e}}, {Shah}, {Shen}, {Simons}, {Snyder}, {Stanway},
  {Straughn}, {Teplitz}, {Vanderhoof}, {Vega-Ferrero}, {Wang}, {Weiner},
  {Willmer}, {Wuyts}, \& {Zavala}}]{Finkelstein2022}
{Finkelstein}, S.~L., {Bagley}, M.~B., {Arrabal Haro}, P., {et~al.} 2022, arXiv
  e-prints, arXiv:2207.12474

\bibitem[{{Fischer} {et~al.}(2014){Fischer}, {Abel}, {Gonz{\'a}lez-Alfonso},
  {Dudley}, {Satyapal}, \& {van Hoof}}]{Fischer2014}
{Fischer}, J., {Abel}, N.~P., {Gonz{\'a}lez-Alfonso}, E., {et~al.} 2014, \apj,
  795, 117

\bibitem[{{Gallazzi} {et~al.}(2005){Gallazzi}, {Charlot}, {Brinchmann},
  {White}, \& {Tremonti}}]{Gallazzi2005}
{Gallazzi}, A., {Charlot}, S., {Brinchmann}, J., {White}, S. D.~M., \&
  {Tremonti}, C.~A. 2005, \mnras, 362, 41

\bibitem[{{Gururajan} {et~al.}(2021){Gururajan}, {B{\'e}thermin}, {Theul{\'e}},
  {Spilker}, {Aravena}, {Archipley}, {Chapman}, {DeBreuck}, {Gonzalez},
  {Hayward}, {Hezaveh}, {Hill}, {Jarugula}, {Litke}, {Malkan}, {Marrone},
  {Narayanan}, {Phadke}, {Reuter}, {Vieira}, {Vizgan}, \&
  {Wei{\ss}}}]{Gururajan2021}
{Gururajan}, G., {B{\'e}thermin}, M., {Theul{\'e}}, P., {et~al.} 2021, arXiv
  e-prints, arXiv:2109.03450

\bibitem[{{Harikane} {et~al.}(2020){Harikane}, {Ouchi}, {Inoue}, {Matsuoka},
  {Tamura}, {Bakx}, {Fujimoto}, {Moriwaki}, {Ono}, {Nagao}, {Tadaki}, {Kojima},
  {Shibuya}, {Egami}, {Ferrara}, {Gallerani}, {Hashimoto}, {Kohno}, {Matsuda},
  {Matsuo}, {Pallottini}, {Sugahara}, \& {Vallini}}]{Harikane2020}
{Harikane}, Y., {Ouchi}, M., {Inoue}, A.~K., {et~al.} 2020, \apj, 896, 93

\bibitem[{{Herrera-Camus} {et~al.}(2018){Herrera-Camus}, {Sturm},
  {Graci{\'a}-Carpio}, {Lutz}, {Contursi}, {Veilleux}, {Fischer},
  {Gonz{\'a}lez-Alfonso}, {Poglitsch}, {Tacconi}, {Genzel}, {Maiolino},
  {Sternberg}, {Davies}, \& {Verma}}]{Herrera-Camus2018}
{Herrera-Camus}, R., {Sturm}, E., {Graci{\'a}-Carpio}, J., {et~al.} 2018, \apj,
  861, 95

\bibitem[{{Kelly}(2007)}]{Kelly2007}
{Kelly}, B.~C. 2007, \apj, 665, 1489

\bibitem[{{Kewley} {et~al.}(2019){Kewley}, {Nicholls}, \&
  {Sutherland}}]{Kewley2019}
{Kewley}, L.~J., {Nicholls}, D.~C., \& {Sutherland}, R.~S. 2019, \araa, 57, 511

\bibitem[{{Kojima} {et~al.}(2017){Kojima}, {Ouchi}, {Nakajima}, {Shibuya},
  {Harikane}, \& {Ono}}]{Kojima2017}
{Kojima}, T., {Ouchi}, M., {Nakajima}, K., {et~al.} 2017, \pasj, 69, 44

\bibitem[{{Komatsu} {et~al.}(2011){Komatsu}, {Smith}, {Dunkley}, {Bennett},
  {Gold}, {Hinshaw}, {Jarosik}, {Larson}, {Nolta}, {Page}, {Spergel},
  {Halpern}, {Hill}, {Kogut}, {Limon}, {Meyer}, {Odegard}, {Tucker}, {Weiland},
  {Wollack}, \& {Wright}}]{Komatsu2011}
{Komatsu}, E., {Smith}, K.~M., {Dunkley}, J., {et~al.} 2011, \apjs, 192, 18

\bibitem[{{Lagache} {et~al.}(2018){Lagache}, {Cousin}, \&
  {Chatzikos}}]{Lagache2018}
{Lagache}, G., {Cousin}, M., \& {Chatzikos}, M. 2018, \aap, 609, A130

\bibitem[{{Luhman} {et~al.}(2003){Luhman}, {Satyapal}, {Fischer}, {Wolfire},
  {Sturm}, {Dudley}, {Lutz}, \& {Genzel}}]{Luhman2003}
{Luhman}, M.~L., {Satyapal}, S., {Fischer}, J., {et~al.} 2003, \apj, 594, 758

\bibitem[{{Ma} {et~al.}(2016){Ma}, {Gonzalez}, {Vieira}, {Aravena}, {Ashby},
  {B{\'e}thermin}, {Bothwell}, {Brandt}, {de Breuck}, {Carlstrom}, {Chapman},
  {Gullberg}, {Hezaveh}, {Litke}, {Malkan}, {Marrone}, {McDonald}, {Murphy},
  {Spilker}, {Sreevani}, {Stark}, {Strandet}, \& {Wang}}]{Ma2016}
{Ma}, J., {Gonzalez}, A.~H., {Vieira}, J.~D., {et~al.} 2016, \apj, 832, 114

\bibitem[{{Maiolino} \& {Mannucci}(2019)}]{Maiolino2019}
{Maiolino}, R. \& {Mannucci}, F. 2019, \aapr, 27, 3

\bibitem[{{Noll} {et~al.}(2009){Noll}, {Burgarella}, {Giovannoli}, {Buat},
  {Marcillac}, \& {Mu{\~n}oz-Mateos}}]{Noll2009}
{Noll}, S., {Burgarella}, D., {Giovannoli}, E., {et~al.} 2009, \aap, 507, 1793

\bibitem[{{Onodera} {et~al.}(2016){Onodera}, {Carollo}, {Lilly}, {Renzini},
  {Arimoto}, {Capak}, {Daddi}, {Scoville}, {Tacchella}, {Tatehora}, \&
  {Zamorani}}]{Onodera2016}
{Onodera}, M., {Carollo}, C.~M., {Lilly}, S., {et~al.} 2016, \apj, 822, 42

\bibitem[{{Peeples} {et~al.}(2008){Peeples}, {Pogge}, \&
  {Stanek}}]{Peeples2008}
{Peeples}, M.~S., {Pogge}, R.~W., \& {Stanek}, K.~Z. 2008, \apj, 685, 904

\bibitem[{{Reuter} {et~al.}(2020){Reuter}, {Vieira}, {Spilker}, {Weiss},
  {Aravena}, {Archipley}, {B{\'e}thermin}, {Chapman}, {De Breuck}, {Dong},
  {Everett}, {Fu}, {Greve}, {Hayward}, {Hill}, {Hezaveh}, {Jarugula}, {Litke},
  {Malkan}, {Marrone}, {Narayanan}, {Phadke}, {Stark}, \&
  {Strandet}}]{Reuter2020}
{Reuter}, C., {Vieira}, J.~D., {Spilker}, J.~S., {et~al.} 2020, \apj, 902, 78

\bibitem[{{Roberts-Borsani} {et~al.}(2020){Roberts-Borsani}, {Ellis}, \&
  {Laporte}}]{Roberts-Borsani2020}
{Roberts-Borsani}, G.~W., {Ellis}, R.~S., \& {Laporte}, N. 2020, \mnras, 497,
  3440

\bibitem[{{R{\"o}llig} {et~al.}(2007){R{\"o}llig}, {Abel}, {Bell}, {Bensch},
  {Black}, {Ferland}, {Jonkheid}, {Kamp}, {Kaufman}, {Le Bourlot}, {Le Petit},
  {Meijerink}, {Morata}, {Ossenkopf}, {Roueff}, {Shaw}, {Spaans}, {Sternberg},
  {Stutzki}, {Thi}, {van Dishoeck}, {van Hoof}, {Viti}, \&
  {Wolfire}}]{RolligAA07}
{R{\"o}llig}, M., {Abel}, N.~P., {Bell}, T., {et~al.} 2007, \aap, 467, 187

\bibitem[{{Schaerer} {et~al.}(2022){Schaerer}, {Marques-Chaves}, {Barrufet},
  {Oesch}, {Izotov}, {Naidu}, {Guseva}, \& {Brammer}}]{Schaerer2022}
{Schaerer}, D., {Marques-Chaves}, R., {Barrufet}, L., {et~al.} 2022, arXiv
  e-prints, arXiv:2207.10034

\bibitem[{{Seaquist} {et~al.}(2004){Seaquist}, {Yao}, {Dunne}, \&
  {Cameron}}]{Seaquist2004}
{Seaquist}, E., {Yao}, L., {Dunne}, L., \& {Cameron}, H. 2004, \mnras, 349,
  1428

\bibitem[{{Seymour} {et~al.}(2012){Seymour}, {Altieri}, {De Breuck}, {Barthel},
  {Coia}, {Conversi}, {Dannerbauer}, {Dey}, {Dickinson}, {Drouart}, {Galametz},
  {Greve}, {Haas}, {Hatch}, {Ibar}, {Ivison}, {Jarvis}, {Kov{\'a}cs}, {Kurk},
  {Lehnert}, {Miley}, {Nesvadba}, {Rawlings}, {Rettura}, {R{\"o}ttgering},
  {Rocca-Volmerange}, {S{\'a}nchez-Portal}, {Santos}, {Stern}, {Stevens},
  {Valtchanov}, {Vernet}, \& {Wylezalek}}]{Seymour2012}
{Seymour}, N., {Altieri}, B., {De Breuck}, C., {et~al.} 2012, \apj, 755, 146

\bibitem[{{Smail} {et~al.}(2011){Smail}, {Swinbank}, {Ivison}, \&
  {Ibar}}]{Smail2011}
{Smail}, I., {Swinbank}, A.~M., {Ivison}, R.~J., \& {Ibar}, E. 2011, \mnras,
  414, L95

\bibitem[{{Snijders} {et~al.}(2007){Snijders}, {Kewley}, \& {van der
  Werf}}]{Snijders2007}
{Snijders}, L., {Kewley}, L.~J., \& {van der Werf}, P.~P. 2007, \apj, 669, 269

\bibitem[{{Spilker} {et~al.}(2018){Spilker}, {Aravena}, {B{\'e}thermin},
  {Chapman}, {Chen}, {Cunningham}, {De Breuck}, {Dong}, {Gonzalez}, {Hayward},
  {Hezaveh}, {Litke}, {Ma}, {Malkan}, {Marrone}, {Miller}, {Morningstar},
  {Narayanan}, {Phadke}, {Sreevani}, {Stark}, {Vieira}, \&
  {Wei{\ss}}}]{Spilker2018}
{Spilker}, J.~S., {Aravena}, M., {B{\'e}thermin}, M., {et~al.} 2018, Science,
  361, 1016

\bibitem[{{Spinoglio} {et~al.}(2015){Spinoglio}, {Pereira-Santaella}, {Dasyra},
  {Calzoletti}, {Malkan}, {Tommasin}, \& {Busquet}}]{Spinoglio2015}
{Spinoglio}, L., {Pereira-Santaella}, M., {Dasyra}, K.~M., {et~al.} 2015, \apj,
  799, 21

\bibitem[{{Strandet} {et~al.}(2016){Strandet}, {Weiss}, {Vieira}, {de Breuck},
  {Aguirre}, {Aravena}, {Ashby}, {B{\'e}thermin}, {Bradford}, {Carlstrom},
  {Chapman}, {Crawford}, {Everett}, {Fassnacht}, {Furstenau}, {Gonzalez},
  {Greve}, {Gullberg}, {Hezaveh}, {Kamenetzky}, {Litke}, {Ma}, {Malkan},
  {Marrone}, {Menten}, {Murphy}, {Nadolski}, {Rotermund}, {Spilker}, {Stark},
  \& {Welikala}}]{Strandet2016}
{Strandet}, M.~L., {Weiss}, A., {Vieira}, J.~D., {et~al.} 2016, \apj, 822, 80

\bibitem[{{Taylor} {et~al.}(2022){Taylor}, {Barger}, \& {Cowie}}]{Taylor2022}
{Taylor}, A.~J., {Barger}, A.~J., \& {Cowie}, L.~L. 2022, arXiv e-prints,
  arXiv:2208.06418

\bibitem[{{Trump} {et~al.}(2022){Trump}, {Arrabal Haro}, {Simons}, {Backhaus},
  {Amor{\'\i}n}, {Dickinson}, {Fern{\'a}ndez}, {Papovich}, {Nicholls},
  {Kewley}, {Brunker}, {Salzer}, {Wilkins}, {Almaini}, {Bagley}, {Berg},
  {Bhatawdekar}, {Bisigello}, {Buat}, {Burgarella}, {Calabr{\`o}}, {Casey},
  {Ciesla}, {Cleri}, {Cole}, {Cooper}, {Cooray}, {Costantin}, {Ferguson},
  {Finkelstein}, {Fujimoto}, {Gardner}, {Gawiser}, {Giavalisco}, {Grazian},
  {Grogin}, {Hathi}, {Hirschmann}, {Holwerda}, {Huertas-Company}, {Hutchison},
  {Jogee}, {Juneau}, {Jung}, {Kartaltepe}, {Kirkpatrick}, {Koekemoer}, {Lotz},
  {Lucas}, {Magnelli}, {Matharu}, {P{\'e}rez-Gonz{\'a}lez}, {Pirzkal},
  {Rafelski}, {Rose}, {Seill{\'e}}, {Somerville}, {Straughn}, {Tacchella},
  {Vanderhoof}, {Weiner}, {Wuyts}, {Yung}, \& {Zavala}}]{Trump2022}
{Trump}, J.~R., {Arrabal Haro}, P., {Simons}, R.~C., {et~al.} 2022, arXiv
  e-prints, arXiv:2207.12388

\bibitem[{{Yeh} \& {Matzner}(2012)}]{Yeh2012}
{Yeh}, S. C.~C. \& {Matzner}, C.~D. 2012, \apj, 757, 108

\end{thebibliography}

\appendix

\section{Contribution of emission lines to broad bands}
  \label{Annex.line_contributions}

In this appendix, we quantitatively estimate what the contribution of the far-IR emission lines to the broad bands is, and we show that only the emission lines with the largest equivalent widths, namely [OIII]88.3 \si{\um} and [CII]157.6 \si{\um} for the considered broad bands and redshift range could have an impact on the observed colours.

The offset in the colour-colour diagramme (Fig.~\ref{Fig.CIGALE_color_models}) could be due to changes of the PSW$_{250 \si{\um}}$ / PMW$_{350 \si{\um}}$ and/or the LABOCA$_{870 \si{\um}}$ / PLW$_{500 \si{\um}}$ colours. Even though this is an over simplification (Fig.~\ref{Fig.sample_spectra_redshifts}), for the sake of clarity, we assume that the main effect is due to [CII]157.6 \si{\um} entering or exiting from the LABOCA$_{870 \si{\um}}$, and nothing else is modified. The flux density in band PLW$_{500 \si{\um}}$ does not change over the redshift range 4 $\lesssim$ z $\lesssim$ 6. In this case, we have:

\begin{figure*}
  \centering
  \includegraphics[width=0.62\textwidth, angle=0]{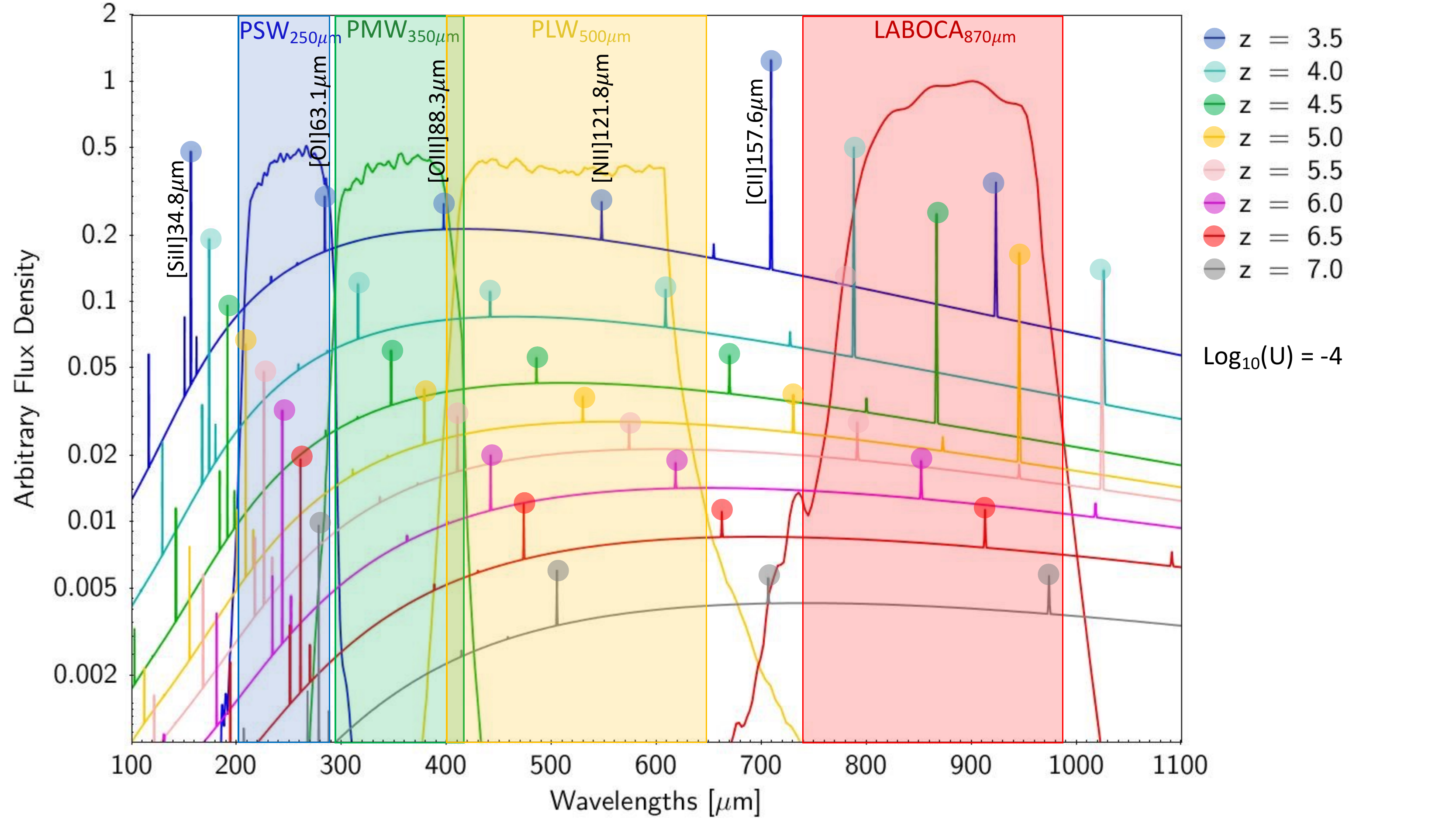}
  \includegraphics[width=0.62\textwidth, angle=0]{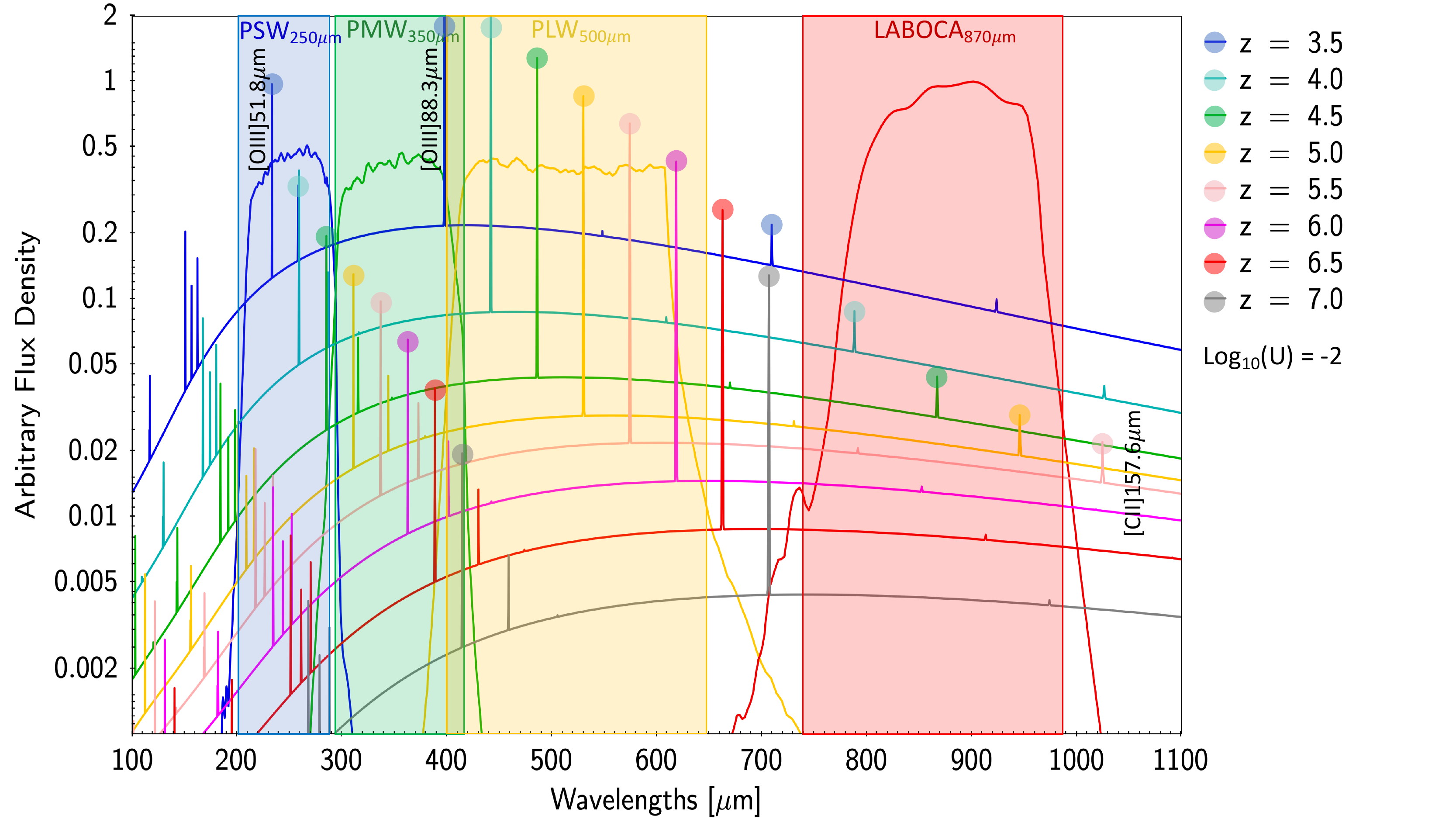}
  \caption{Top: At log$_{10}$ U = -4.0, the strong [CII]157.6 \si{\um} lines enters into the LABOCA$_{870 \si{\um}}$ filters at z $\sim$ 4, and exits at z $\sim$ 5.2. The line induces an upward move of the corresponding $\log_{10} \frac{LABOCA_{870  \si{\um}}} {PLW_{500 \si{\um}}}$ colour. [SIII]34.8 \si{\um} enters into the PSW$_{250 \si{\um}}$ at z $\sim$ = 5.5 which boosts the $\log_{10} \frac{PSW_{250  \si{\um}}} {PMW_{350 \si{\um}}}$ colour. We find a total move of the galaxies by about $\Delta_{total} \sim 0.10$ in the colour-colour diagramme. These two combined effects explains that PDR-dominated galaxies at 4 $\lesssim$ z $\lesssim$ 6 are offset from the other galaxies. Bottom: the move of the [OIII]51.8 \si{\um} and [OIII]88.3 \si{\um} lines of galaxies with a strong emission from HII regions at log$_{10}$ U = -2.0 could induce specific colours that could help identifying such galaxies from this colour-colour diagramme. However, the effect is less clear as models with no lines can also lie here (Fig.~\ref{Fig.CIGALE_color_models}).  }
  \label{Fig.sample_spectra_redshifts}
\end{figure*}

\begin{equation}
$$F_{PLW\_500 \si{\um}}^{line} = F_{PLW\_500 \si{\um}}^{noline} ~\&~ F_{LABOCA\_870 \si{\um}}^{line} \gtrsim F_{PLW\_500 \si{\um}}^{noline}$$. 
\end{equation}

We define the band ratio: 
\begin{equation}
$$R^{line}_{870 \si{\um}\_ 500 \si{\um}} = \frac{F_{LABOCA\_870 \si{\um}}^{line} \Delta \lambda_{870 \si{\um}}}            
                                                                           {F_{PLW\_500 \si{\um}}^{line} \Delta \lambda_{500 \si{\um}}}$$ 
\end{equation}

where $\Delta \lambda_{500 \si{\um}} = 143~\si{\um}$, $\Delta \lambda_{500 \si{\um}} = 186~\si{\um}$, $\Delta \lambda_{250 \si{\um}} = 67~\si{\um}$, and $\Delta \lambda_{350 \si{\um}} = 95~\si{\um}$ (SVO Filter Service). Using the right-end part of Eq.~C.1, we have: 

\begin{equation}
$$R^{line}_{870 \si{\um}\_ 500 \si{\um}} \gtrsim R^{noline}_{870 \si{\um}\_ 500 \si{\um}}$$
\end{equation}

From Eq.~C.2, where F$_{line}$ is the line flux contributing to the LABOCA$_{870 \si{\um}}$ band:

\begin{equation}
$$R^{line}_{870 \si{\um}\_ 500 \si{\um}} = \frac{F_{LABOCA\_870 \si{\um}}^{noline} \Delta \lambda_{870 \si{\um}} + F_{line}} {F_{PLW\_500 \si{\um}}^{noline} \Delta \lambda_{500 \si{\um}}} $$
\end{equation}

From the left-end part of Eq.~C.1, and the definition of the equivalent width (EW$_{line}$), we have  F$_{line}$ =  F$_{LABOCA\_870 \si{\um}}^{noline}$ EW$_{line}$. This gives:

\begin{equation}
$$R^{line}_{870 \si{\um}\_ 500 \si{\um}} = \frac{F_{LABOCA\_870 \si{\um}}^{noline}} {F_{PLW\_500 \si{\um}}^{noline} \Delta \lambda_{500 \si{\um}}} (\Delta \lambda_{870 \si{\um}} + EW_{line}) $$
\end{equation}

And finally:

\begin{equation}
$$R^{line}_{870 \si{\um}\_ 500 \si{\um}} = R^{noline}_{870 \si{um}\_ 500 \si{\um}}\ (1 + \frac{EW_{line}}{\Delta \lambda_{870 \si{\um}}}) $$
\end{equation}
The mean colours of the outliers are $\langle \log_{10}$ (PSW$_{250 \si{\um}}$ / PMW$_{350 \si{\um}}) \rangle$ = -0.205 $\pm$ 0.113 and $\langle \log_{10}$(LABOCA$_{870 \si{\um}}$ / PLW$_{500 \si{um}} \rangle$ = -0.187 $\pm$ 0.113. For the same redshift range, CIGALE models without line gives $\langle \log_{10}$ (PSW$_{250 \si{\um}}$ / PMW$_{350 \si{\um}}) \rangle$ = -0.254 $\pm$ 0.134 and $\langle \log_{10}$(LABOCA$_{870 \si{\um}}$ / PLW$_{500 \si{\um}} \rangle$ = -0.101 $\pm$ 0.194 The average move needed to reach these galaxies amounts to $\mid \Delta (\log_{10} (PSW_{250 \si{\um}} / PMW_{350 \si{\um}}) \mid $ $\approx$ 0.05 and $\mid \Delta (\log_{10}(LABOCA_{870 \si{\um}} / PLW_{500 \si{\um}}) \mid $ $\approx$ 0.09. That is a total offset of the order of $\Delta_{total} \sim 0.10$. Because we cannot know where the galaxy would be, without accounting for the emission lines, both colour could contribute to this offset, but it is likely that [CII]157.6 \si{\um} is dominant.

We find that the models inside the ellipses, accounting for the observed uncertainties, have 0.6 $\lesssim$ EW([CII]157.6 \si{\um}) $\lesssim$ 25.0 \si{\um}. From this, we get:

\begin{equation}
$$1.0 \lesssim \frac {R^{line}_{870 \si{\um}\_ 500 \si{\um}}}{R^{noline}_{870 \si{\um}\_ 500 \si{\um}}} \lesssim 1.2$$
\end{equation}
\begin{equation}
$$0.0 \lesssim \Delta (\log_{10}(LABOCA_{870 \si{\um}} / PLW_{500 \si{\um}}) \lesssim 0.1$$
\end{equation}
that is an offset in the colour-colour diagramme $\lesssim$ 0.1 which is about the same order or the maximum one observed.

The [SIII]33.47 \si{um} line enters into the PSW$_{250 \si{\um}}$ at about the same redshift range and modifies $\log_{10}$ (PSW$_{250 \si{\um}}$ / PMW$_{350 \si{\um}}$). However, with 0.05 $\lesssim$ EW([SIII]33.5 \si{\um}) $\lesssim$ 0.26, we estimate that the [SiII]33.47 \si{\um} line should not significantly contribute to the PSW$_{250 \si{\um}}$ band. 

No CIGALE models at $\delta z=\pm0.5$ match the two top-most objects in Fig.~\ref{Fig.CIGALE_color_models}: SPT0245-63 at z = 5.626 and SPT0243-49 at z=5.702, and we cannot estimate any of the physical parameters. The reasons why we cannot reproduce the colours of these two objects is uncertain. Two plausible hypotheses could be made, though. First, the flux densities of these objects might not be correct or, at least, the uncertainties could be under estimated. The other explanation might be that our present grid of models might not cover the entire possible range of data. These two objects would deserve a closer analysis.


\end{document}